\documentclass[%
 reprint,
 amsmath,amssymb,
 aps,
pre,
]{revtex4-2}

\DeclareUnicodeCharacter{0229}{\c{e}}
\usepackage{graphicx}
\usepackage{dcolumn}
\usepackage{bm}

\begin{document}

\preprint{Preprint ID}

\title{Thermodynamic control of non-equilibrium systems}

\author{Dana Kamp}
\email{dana.taylor.kamp@nbi.ku.dk}

\author{Karel Proesmans}
\email{karel.proesmans@nbi.ku.dk}
\affiliation{Niels Bohr International Academy, Niels Bohr Institute, University of Copenhagen, Blegdamsvej 17, 2100 Copenhagen, Denmark}
\vspace{10pt}

\date{\today}

\begin{abstract}
We study the thermodynamic cost associated with driving systems between different non-equilibrium steady states. In particular, we combine a linear-response framework for non-equilibrium Markov systems with Lagrangian techniques to minimize the dissipation associated with driving processes. We then apply our framework to a simple toy model. Our results show several remarkable properties for the optimal protocol, such as diverging parameters and finite entropy production in the slow-driving limit.
\end{abstract}


\maketitle


\section{\label{sec:introduction} Introduction}
Thermodynamic optimization of finite-time protocols plays an important role in a broad range of fields. One can for example think about the erasure of a bit \cite{aurell2012refined,zulkowski2014optimal,proesmans_finite-time_2020,zhen2021universal,basile2024learning}, shortcuts to adiabaticity \cite{tu2014stochastic,takahashi2017shortcuts,guery2019shortcuts} and the driving of molecular motors \cite{gupta2022optimal,brown2019pulling,lucero2019optimal}.

Based on the framework of stochastic thermodynamics, a number of methods have been developed to study this problem. One prominent method is to use mathematical optimal transport theory to find protocols that minimise entropy production \cite{aurell2011optimal,van2023thermodynamic}. The main drawbacks of this method are that one generally needs to assume full control of the energy landscape and that it becomes computationally unfeasible for high-dimensional systems. These problems are circumvented by the method developed in \cite{crooks_measuring_2007, sivak_thermodynamic_2012}. Here, one uses linear response theory around equilibrium to develop a geometric picture to calculate the entropy production associated with systems with parametric control. This can then be used to minimize the entropy production in a computationally efficient way. This method can be used for slowly-driven systems that are always kept close to equilibrium.

Many realistic systems do, however, operate far from thermodynamic equilibrium. For example, cellular biological processes generally operate in a non-equilibrium environment \cite{ritort2008nonequilibrium, fang2019nonequilibrium}, and microelectronic memories constantly dissipate energy \cite{freitas2021stochastic,freitas_reliability_2022}. This makes it impossible to apply the afore-mentioned geometric framework directly to these systems. This has lead to a number of extensions of the original framework. Firstly, it can be shown that applying the framework directly to non-equilibrium systems leads to an optimization of the non-adiabatic part of the entropy production, rather than the full entropy production \cite{mandal_analysis_2016}. Furthermore, extensions of the original framework have been made for the specific class of active matter systems \cite{davis_active_2024}. However, to the best of our knowledge, no linear-repsonse framework exists to minimize the entropy production of general far-from-equilibrium systems. In this paper, we close this gap by calculating a general formula for the entropy production of slowly driven systems far from thermodynamic equilibrium. This leads to a similar optimization scheme to the one developed in \cite{crooks_measuring_2007, sivak_thermodynamic_2012}  for equilibrium systems.

In the next section we will discuss the theoretical background and develop the general linear response framework. Subsequently, we will look at the optimization of the entropy production in section \ref{sec:optimization}. In section \ref{sec:three_state_system}, we will apply it to a simple toy model where we will also discuss the main differences and similarities with the near-equilibriium framework. Finally, in section \ref{sec:discussion}, we will discuss the main results and give a perspective for future research.

\section{\label{sec:theory_basic} Stochastic thermodynamics and entropy production} 
   We are working with a Markov system that can exist in several discrete states $i \in \{1,.., N\}$ with probabilities $p_i(t)$. How the probability of being in a specific state, $i$, changes over time, depends on the rates at which the system is moving into, $W_{ij}$ and out of that state, $W_{ji}$, and the distribution of states, $p$:
\begin{eqnarray}
	\frac{d p_{i}}{d t} = \sum _{j \neq i} W_{ij} p_j - \sum _{j \neq i} W_{ji} p_i,
\end{eqnarray}
for each state $i$, or in vectorial form:
\begin{eqnarray}
	\frac{d \mathbf{p}(t)}{dt} =  W \mathbf{p}(t) \label{eq:master_eq}
\end{eqnarray}
with $W_{ii}=-\sum_jW_{ji}$.
This is called the master equation and if all rates $W_{ij}$ are kept constant, the system evolves to a steady-state $\mathbf{ p}_{\mathrm{SS}}$,
\begin{eqnarray}
	\frac{d \mathbf{ p}_{\mathrm{SS}}}{dt} =W\mathbf{ p}_{\mathrm{SS}}= 0.
\end{eqnarray} 

A special case of a steady state is equilibrium , where the probability of moving from state $i$ to $j$ is equal to moving in the opposite direction, 
\begin{eqnarray}W_{ij}p_j = W_{ji}p_i,\label{eq:db}\end{eqnarray} 
a property known as detailed balance. In this case, there are no net flows or cycles between states, and no work can be done.

Often a system is in contact with multiple reservoirs (such as particle and heat reservoir), each of which can separately induce transitions inside the system. The transition rates associated with reservoir $\nu$ are then written as $W^{(\nu)}_{ij}$ and as these reservoirs are assumed to interact independently, and the master equation becomes
\begin{eqnarray}
    W_{ij}=\sum_{\nu}W^{(\nu)}_{ij}.
\end{eqnarray}
In such a case the steady state generally does not satisfy the detailed balance condition, Eq.~\eqref{eq:db}.

The central goal of this paper is to calculate the entropy production of time-dependent systems, that is systems where the transition rates $W_{ij}$ are time-dependent. For systems described by master equations, this entropy production rate is given by \cite{van2015ensemble}:
\begin{eqnarray}
	\sigma(t) = k_B \sum_{i,j,(\nu)} W_{ij}^{(\nu)}(t)p_j\ln\frac{W_{ij}^{(\nu)}(t)p_j(t)}{W_{ji}^{(\nu)(t)}p_i(t)}.\label{eq:entr_prod}
\end{eqnarray}
We will assume that the time-dependent dynamics are driven by a parameter, $\alpha(t)$. This parameter can for example control an energy-level of a state $i$, the temperature of the environment, or the chemical potential of some species. 

\subsection{\label{sec:theory_near_eq_sys} Near-equilibrium systems} 
We want to calculate the entropy production, $\Delta S_{\textrm{tot}}=\int^{t_f}_0dt\,\sigma(t)$ to go from an initial parameter value $\alpha=\alpha_i$ to a final value $\alpha=\alpha_f$ over a duration $t_f$. It is generally impossible to calculate the entropy production of such time-dependent processes analytically, as $\mathbf{p}(t)$ depends on $\alpha(t)$ in a highly non-trivial way, but several methods have been developed to approximate the total entropy production.

One of the most influential methods to do this was developed by Sivak and Crooks \cite{sivak_thermodynamic_2012}. Essentially, the main idea is to assumptions are that there exists an equilbirium probability distribution $p^{\textrm{eq}}$ at all times that satisfies the detailed balance condition, Eq.~\eqref{eq:db}, and that the system is driven sufficiently slowly. Under these circumstances, one can use linear response theory to get an approximation for the entropy production:
\begin{eqnarray}
   \Delta S_{\textrm{tot}} =\int^{t_f}_0dt\,A(\alpha(t))\dot{\alpha}(t)^2,
\end{eqnarray}
where
\begin{eqnarray}
    A(\alpha)=\int^{\infty}_0dt\,\left\langle\frac{\partial X(t)}{\partial \alpha}\frac{\partial X(0)}{\partial \alpha}\right\rangle_{\alpha},\label{eq:Aeq}
\end{eqnarray}
where $\langle\cdot\rangle_{\alpha}$ is the equilibrium steady-state average taken over the dynamics with $\alpha$ fixed at its current value, and
\begin{eqnarray}
    X_i=\ln p^{\textrm{eq}}_i.
\end{eqnarray}
This allows to calculate the entropy production of a broad range of slowly driven systems \cite{blaber2023optimal}, by first numerically calculating $A(\alpha)$ for all values of $\alpha$. One can then use this to derive the optimal protocol, $\{\alpha(t)\}_0^{t_f}$ that minimises $\Delta S_{\textrm{tot}}$. Furthermore, one can use it to show a number of general properties, such as that $\Delta S_{\textrm{tot}}\sim t_f^{-1}$ and  that the protocol that minimises $\Delta S_{\textrm{tot}}$ is independent of the duration of the protocol \cite{deffner2020thermodynamic,blaber2023optimal}.

\subsection{\label{sec:theory_general non_eq_sys} General non-equilibrium systems} 
Several extensions to the near-equilibrium framework from the previous sub-section have been derived. These extensions include active systems \cite{davis_active_2024}, and systems that only take into account part of the entropy production \cite{mandal_analysis_2016}. Despite these advancements, there is, to the best of our knowledge, no general framework to calculate the entropy production of slowly-driven non-equilibrium systems is available. We will develop such a framework in this section.

We will start by expanding the probability distribution $\textbf{p}(t)$ as a function of $\dot{\alpha}(t)$. Following \cite{mandal_analysis_2016}, we first note that (see also appendix \ref{app:expand_state_vec}):
\begin{eqnarray}
	\textbf{p}(t) &=& \sum_{n=0}^{\infty}\left(W^\dagger\frac{d}{dt}\right)^n \textbf{p}_{\mathrm{SS}}(t)\nonumber\\
	&=&\sum_{n=0}^{\infty}\left(W^\dagger\left[\frac{\partial}{\partial t} + \dot\alpha\frac{\partial}{\partial \alpha}\right]\right)^n \textbf{p}_{\mathrm{SS}}(t)
    \label{eq:p_expansion_inf}
\end{eqnarray}
where $W^\dagger$ is the pseudo-inverse of the transition matrix $W$, defined by:
\begin{eqnarray}
	W = \sum_{i=1}^N \lambda_i \cdot \vert \lambda_i \rangle	\langle \lambda_i \vert \rightarrow W^\dagger = \sum_{i= 2}^N \lambda_i^{-1} \cdot \vert \lambda_i \rangle	\langle \lambda_i \vert
\end{eqnarray}
where $\{\langle \lambda_i \vert\}_{i \leq N}$ and $\{\vert \lambda_i \rangle\}_{i \leq N}$ are the left and right eigenvectors of $W$ associated with eigenvalue $\lambda_i$ and $\lambda_1=0$.
If the driving is sufficiently slow, $\textbf{p}(t)$ is well-approximated by its thirth order expansion:
\begin{eqnarray}
	\mathbf{p}(t) \approx 
	p^{(0)}
	+ \dot\alpha \textbf{p}^{(1)} 
	+ (\dot\alpha)^2 \textbf{p}^{(2)} 
	+ \ddot\alpha \textbf{p}^{(3)}
\end{eqnarray}
with
\begin{eqnarray}
	\textbf{p}^{(0)} &=& \textbf{p}_{\mathrm{SS}}(t)
    \label{eq:p_0}\\
    \textbf{p}^{(1)} &=&W^\dagger\frac{\partial}{\partial \alpha}\textbf{p}_{\mathrm{SS}}(t)\\
   \textbf{p}^{(2)} &=& W^\dagger\left(\frac{\partial W^\dagger}{\partial \alpha} \frac{\partial \textbf{p}_{\mathrm{SS}}(t)}{\partial \alpha} \
      + W^\dagger\frac{\partial^2 \textbf{p}_{\mathrm{SS}}(t)}{\partial \alpha^2}\right) \\\
   \textbf{p}^{(3)} &=&\left(W^\dagger\right)^2\frac{\partial \textbf{p}_{\mathrm{SS}}(t)}{\partial \alpha}
\end{eqnarray}

The advantage of doing this expansion is that the probability distribution now only depends on the instantaneous value of $\alpha(t)$ and its derivatives, and the steady-state distribution, and no longer on the full trajectory history. \\

One can use this expansion to get an approximation for the total entropy production rate $\sigma$. In appendix \ref{app:expansion_terms}, we show that

\begin{eqnarray*}
    \sigma(t) = \,& \sigma_ {SS}& \, + \, \dot\alpha (t) \, F \left ( \alpha(t) \right) \\
    &&\, + \, \dot\alpha (t) \left( A^{(1)} (\alpha (t)) + A^{(2)} (\alpha (t)) \right) \dot\alpha (t) \\
    &&\, + \,\ddot\alpha(t) A^{(3)}(\alpha (t))
    \label{eq:expansion_short_form}
\end{eqnarray*}

with

\begin{eqnarray}
    \sigma_{SS} &=&k_B \sum_{i,j,(\nu)} \ln\frac{W_{ij}^{(\nu)}p_{\mathrm{SS}, j}}{W_{ji}^{(\nu)}p_{\mathrm{SS}, i}}W_{ij}^{(\nu)}p_{\mathrm{SS}, j}\label{eq:expansion_terms1}\\
    F &=& k_B  \sum_{i, j, k, (\nu)} \ln\frac{W_{ij}^{(\nu)}p_{\mathrm{SS}, j}}{W_{ji}^{(\nu)}p_{\mathrm{SS}, i}}W_{ij}^{(\nu)}W_{jk}^\dagger\frac{\partial p_{\mathrm{SS}, k}}{\partial \alpha} \\
	A^{(1)} &=&-k_B \sum_{i,j,(\nu)}p_{\mathrm{SS}, i}\frac{\partial \ln p_{\mathrm{SS}, i}}{\partial \alpha}W_{ij}^\dagger\frac{\partial\ln  p_{\mathrm{SS}, j}}{\partial \alpha} \\
	A^{(2)} &=&k_B \sum_{i,j, k, l,(\nu)}\ln\frac{W_{ij}^{(\nu)}p_{\mathrm{SS}, j}}{W_{ji}^{(\nu)}p_{\mathrm{SS}, i}}W_{ij}^{(\nu)}W_{jk}^\dagger\nonumber\\ &&\qquad\qquad\qquad\qquad\quad\times\frac{\partial}{\partial \alpha}\left( W_{kl}^\dagger \frac{\partial p_{\mathrm{SS}, l}}{\partial \alpha} \right)\\
    A^{(3)} &=&k_B  \sum_{i,j,k, l(\nu)}
	\ln\frac{W_{ij}^{(\nu)}p_{\mathrm{SS}, j}}{W_{ji}^{(\nu)}p_{\mathrm{SS}, i}}W_{ij}^{(\nu)}W^\dagger_{jk}W^\dagger_{kl}\nonumber\\ &&\qquad\qquad\qquad \qquad\qquad\qquad\qquad\times\frac{\partial p_{\mathrm{SS}, l}}{\partial \alpha} \quad
    \label{eq:expansion_terms}
\end{eqnarray}

One can use
\begin{eqnarray}
	\ddot\alpha (t)  A^{(3)}= \frac{d}{d t} \left(\dot\alpha A^{(3)}\right)-(\dot\alpha)^2\frac{\partial}{\partial \alpha}A^{(3)}
\end{eqnarray}

To show that the total entropy production associated with the process is given by
\begin{eqnarray}
	\Delta S_{\mathrm{tot}} &=\int_{t_i}^{t_f}\mathcal{L}(\alpha(t),\dot{\alpha}(t))dt +  \left[ \dot\alpha A^{(3)} \right]_{t_i}^{t_f}\label{eq:stotint}
\end{eqnarray}

with

\begin{eqnarray}
    \mathcal{L}&&(\alpha(t),\dot{\alpha}(t))=\sigma_{SS}(\alpha(t))+F(\alpha(t))\dot{\alpha}\nonumber\\ &&+(A^{(1)}(\alpha(t))+A^{(2)}(\alpha(t))-A^{(3)}(\alpha(t))\dot{\alpha}^2(t).\label{eq:Lres}
\end{eqnarray}

This makes it, in principle, possible to calculate the entropy production for slowly driven far-from-equilibrium systems, but it can become complicated to calculate Eqs.~\eqref{eq:expansion_terms1}-\eqref{eq:expansion_terms} for systems with a large state space. Therefore, it can be useful to rewrite these quantities as correlation functions that can be more easily calculated in simulations or experiments, similar to Eq.~\eqref{eq:Aeq}. To do this, we introduce a number of observables:
\begin{enumerate}
\item The average entropy production associated with a specific state:
\begin{eqnarray}
    O_{1; j} =  \sum_{i, (\nu)} W_{ij}^{(\nu)} \ln \frac{W_{ij}^{(\nu)}p_{SS, j}}{W_{ji}^{(\nu)}p_{SS, i}}
    \label{eq:observable_1}
\end{eqnarray}

\item Pertubation of steady state distribution:
\begin{eqnarray}
    O_{2;j} =  \frac{\partial \ln p_{SS, j}}{\partial \alpha}
    \label{eq:observable_2}
\end{eqnarray}

\item Change in rates:
\begin{eqnarray}
    O_{3;ij} =  \delta(t-\tau_{ij})\frac{\partial \ln(W_{ij})}{\partial \alpha}
    \label{eq:observable_3}
\end{eqnarray}

\end{enumerate}
In terms of these observables, one can write
\begin{eqnarray}
    \sigma_{ss}&=&\left\langle O_{1}\right\rangle_{\alpha}\\
    F&=& -\int_0^\infty \langle O_{1} (\tau)  O_{2} (0) \rangle_{\alpha} dt \\
    A^{(1)} &=& -\int_0^\infty \langle O_{2} (\tau)  O_{2} (0) \rangle_{\alpha} dt\\
    A^{(2)} &=& -\int_0^{\infty} \langle O_{1}(\tau) \tau O_{2}(0)\rangle \langle O_{2} \rangle d\tau \nonumber\\
     && - \int_0^{\infty}\int_0^{\infty}\sum_{(\nu)}  \langle O_{1}(\tau_1 + \tau_2) \tau_1 O_{3}(\tau_2) O_{2}(0) \rangle d\tau_1 d\tau_2 \nonumber\\
    && +   \int_0^\infty \langle O_{1}(\tau) \tau \left( \frac{\partial O_{2}} {\partial \alpha} \vert_{t = 0}+O_{2}(0)^2\right) \rangle d\tau\\
    A^{(3)}&=& \int_0^{\infty}\langle O_1(\tau) \tau O_{2}(0)\rangle d\tau
\end{eqnarray}

A proof of these relations is given in appendix \ref{app:observables}.

One can verify from Eqs.~\eqref{eq:expansion_terms1}-\eqref{eq:expansion_terms}  that every term except for $A^{(1)}$ goes to zero in the near-equilibrium limit, as $O_1\rightarrow 0$, and that $A^{(1)}$ reduces to Eq.~\eqref{eq:Aeq}, showing that our results reduce to the correct expression in the near-equilibrium limit.

\section{ \label{sec:optimization} Protocol optimization} 
We can now use the above expressions for the entropy production to find the protocol that minimizes the dissipation while going from an initial to a final value of $\alpha$ over a given protocol duration.  To do this, we rely on Lagrangian techniques.

In order to do this, we need to consider the boundary contribution of the $A_3$-term. This term does not fit into the Lagrangian, but since it is fixed, it is determined solely by the initial value of $\dot\alpha$ of the protocol, which is fully determined by $\alpha$ and its derivative at the end points. It is then identical for all protocols of the same boundary conditions and can therefore be omitted in the optimization.

The Euler-Lagrange equation states that the protocol $\{\alpha(t)\}$ that minimizes the total entropy production
\begin{eqnarray}
    \frac{\partial \mathcal{L}}{\partial \alpha}( \alpha, 
    \dot\alpha)= 
    \frac{d}{dt}\frac{\partial \mathcal{L}}{\partial \dot\alpha}( \alpha, 
    \dot\alpha),
\end{eqnarray}
This equation can be rewritten as
\begin{eqnarray}
    \ddot\alpha=&& \frac{1}{2(A^{(1)} +A^{(2)}-A^{(3)} )} \nonumber \\&&\quad \times\left[\frac{\partial 
    \sigma^{SS}}{\partial \alpha} -  (\dot\alpha)^2 \frac{\partial 
    }{\partial \alpha}(A^{(1)} +A^{(2)}-A^{(3)} ) \right]
    \label{eq:full_df_eq}
\end{eqnarray}
Note that this equation can be rewritten as
\begin{eqnarray}
    \frac{d}{dt}\left[ (\dot \alpha)^ 2  (A^{(1)} +A^{(2)}-A^{(3)} )-\sigma^{SS}\right]=0,
\end{eqnarray}
or
\begin{eqnarray}
	(\dot\alpha)^2(A^{(1)} + A^{(2)} - A^{(3)}) - \sigma^{SS} = E
\end{eqnarray}
for some constant $E$. This in turn can be written as
\begin{eqnarray}
   \dot \alpha= \pm\sqrt{\frac{E + \sigma^{SS}}{ A^{(1)} +A^{(2)}-A^{(3)} }} \label{eq:alpha_dot_alpha}
\end{eqnarray}
This finally can be integrated to find a relation between the protocol duration and $E$:
\begin{eqnarray}
    t_p = \left|{ \int_{\alpha_0 }^{\alpha_f}} 
    \left(\sqrt{\frac{ A^{(1)} +A^{(2)}-A^{(3)}}{E +\sigma^{SS}}} \right)d\alpha \right|. \label{eq:analytic_tp}
\end{eqnarray}
To calculate the protocol that minimizes dissipation, one can now numerically solve Eq.~\eqref{eq:analytic_tp} to obtain $E$ as a function of $t_p$, and subsequently integrate Eq.~\eqref{eq:alpha_dot_alpha} to get the optimal protocol.

\section{\label{sec:three_state_system} Example: three-state system} 

\begin{figure}
    \centering
    \includegraphics[width=0.5\linewidth]{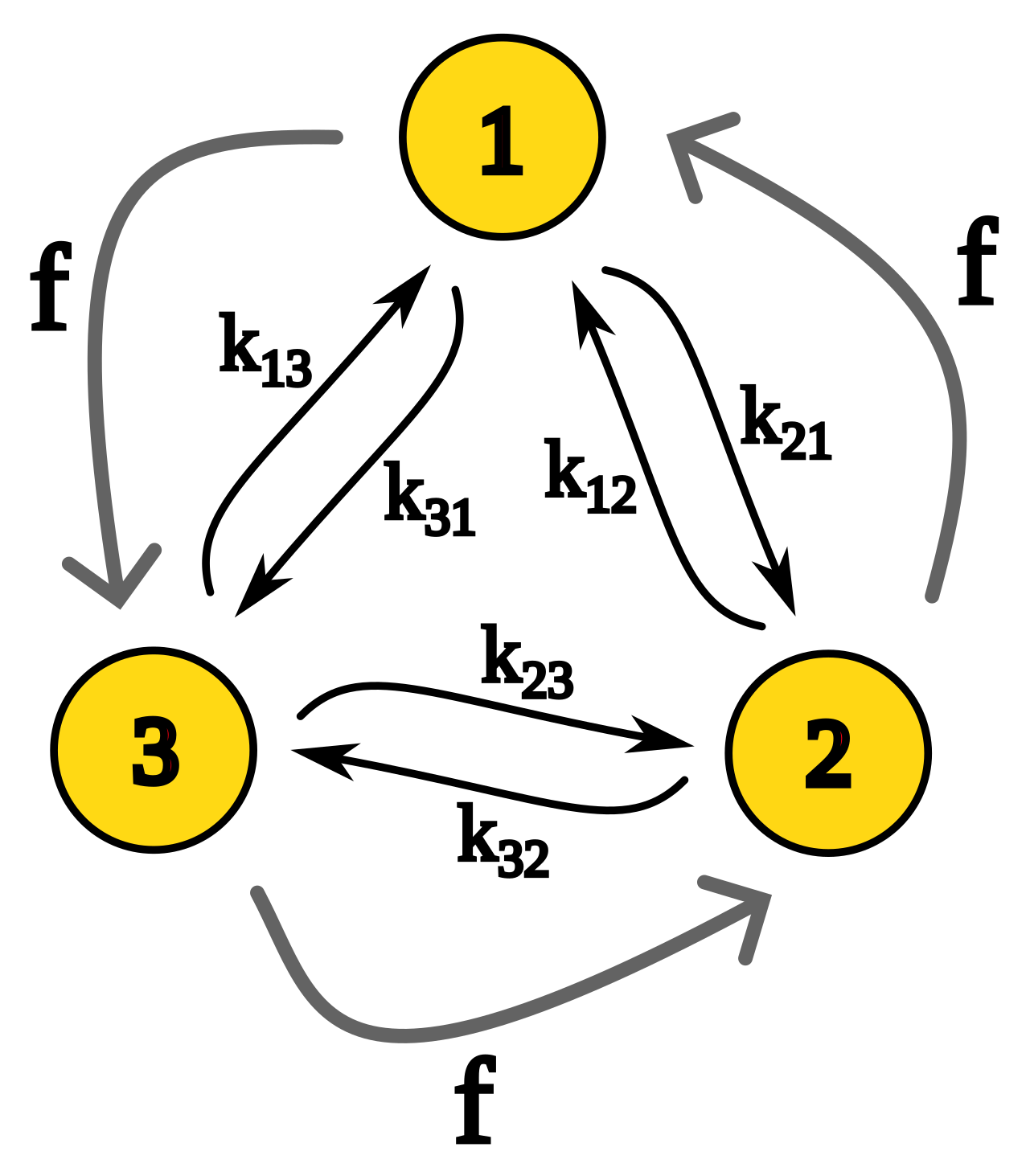}
    \caption{\textbf{The three state system.} The system can exist in 3 different states, depicted by the yellow circles 1, 2 and 3. $k_{ij}$ denotes the rates at which the system changes from one state to the other. $f$ is an externally imposed force that makes sure the system does not relax to equilibrium. }
    \label{fig:three_state_system}
\end{figure}

We now test our framework to a simple toy model, namely discrete three-state system, depicted in Fig.~\ref{fig:three_state_system}. This model has previously been used to model a broad range of biological systems such as cellular ion channels \cite{uteshev_phasic_1996}, and ATP-synthesis \cite{boyer_new_1973, lucero_optimal_2019, large_optimal_2019}.

\begin{figure}
    \centering
    \includegraphics[width=0.5\linewidth]{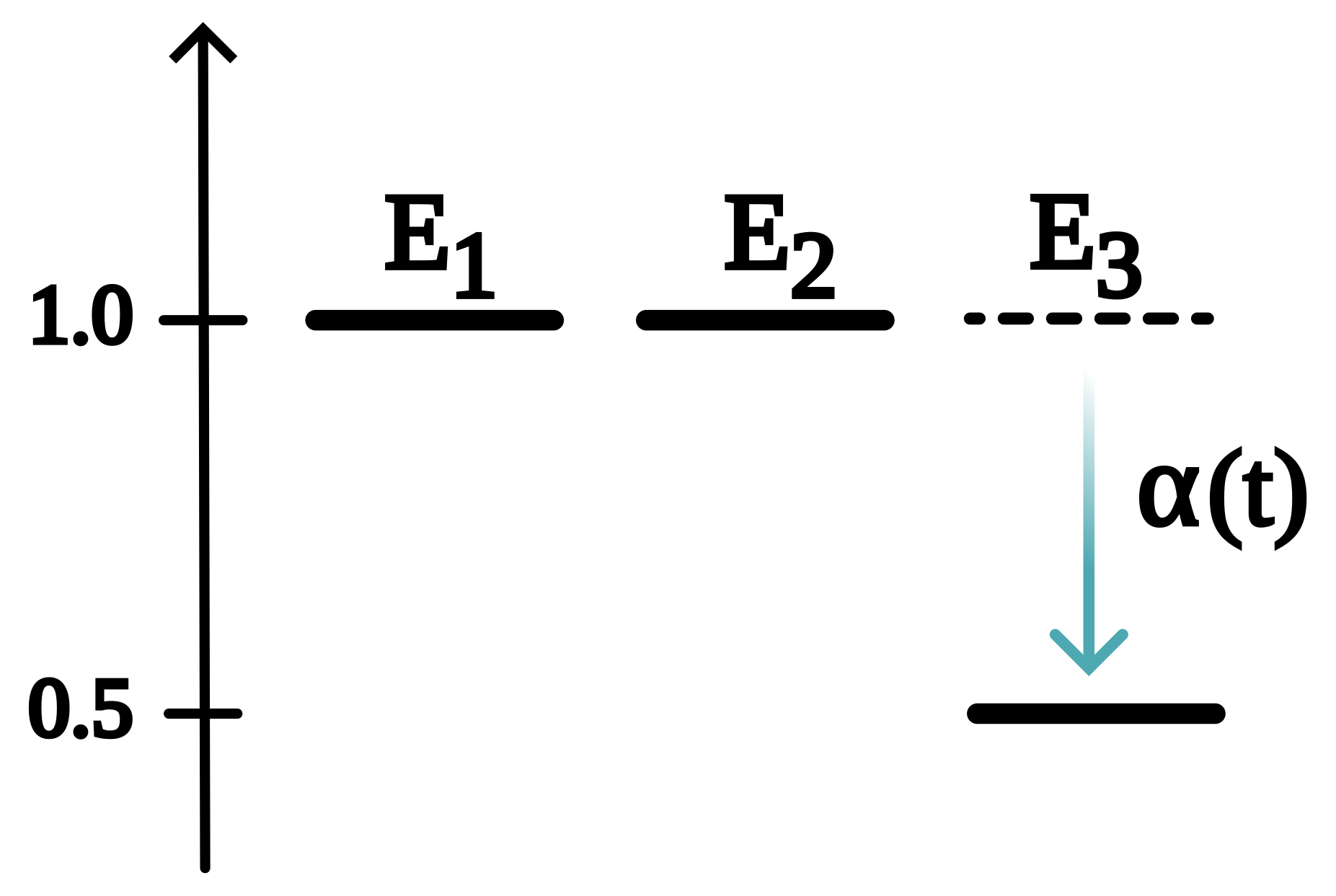}
    \caption{\textbf{Energy levels.} For simplicity, energy levels $E_1$ and $E_2$ are kept constant, while $E_3$ is lowered from $\alpha = k_BT$ to $\alpha = k_BT/2$ through a time-dependent protocol.}
    \label{fig:energy_levels}
\end{figure}

As the control parameter, we will take the energy level of one of the states, namely state $3$, $E_3(t)=\alpha(t)$. Our aim is to minimise the entropy production of driving the system from an initial state state, $E_3(0) = k_BT$ to a final state, $E_3(t_f) =k_BT/2$ through various time-dependent protocols for the parameter $\alpha$, see fig.~\ref{fig:energy_levels}.

The system has the following energy levels:
\begin{eqnarray}
	&&E_1 =k_BT \quad \mathrm{(fixed)}\nonumber\\
	&&E_2 = k_BT\quad \mathrm{(fixed)}\\
	&&E_3 = \alpha(t) \quad \mathrm{where\:} \alpha(0)=k_BT, \, \alpha(t_f)=k_BT/2.\nonumber
\end{eqnarray}
i.e., energy levels $1$ and $2$ are kept constant throughout the protocol.

The jump-rates are modeled as: 

\begin{eqnarray}
	& k_{12} =  \omega_{12} \exp\left[\frac{E_2 + f/2}{k T}\right] \quad
	 k_{21} =  \omega_{21} \exp\left[\frac{E_1 - f/2}{k T}\right] \quad\\\nonumber\\
	& k_{13} =  \omega_{13} \exp\left[\frac{E_3 - f/2}{k T}\right] \quad
	 k_{31} =  \omega_{31} \exp\left[\frac{E_1 + f/2}{k T}\right] \quad \\\nonumber\\
	& k_{23} =  \omega_{23} \exp\left[\frac{E_3 + f/2}{k T}\right] \quad
	 k_{32} =  \omega_{32} \exp\left[\frac{E_2 - f/2}{k T}\right] \quad
\end{eqnarray}
where intrinsic rates $\omega_{ij}=1$ for all $i$ and $j$. One can verify that this system satisfies detailed balance, Eq.~\eqref{eq:db} if $f=0$. As the focus of this paper is on non-equilibrium systems, we break detailed balance by adding a constant non-conservative force, $f = k_BT/10$, to the system. This force pushes a cycle from $ 3 \rightarrow 2 \rightarrow 1$, and leads to steady-state entropy production.

The general transition matrix has the form:
\begin{eqnarray}
W = \begin{bmatrix}
	-(k_{21} + k_{31}) & k_{12} & k_{13} \\
	k_{21} & -(k_{12} + k_{32}) & k_{23} \\
	k_{31} & k_{32} & -(k_{13} + k_{23}) 
\end{bmatrix} 
\end{eqnarray} 
Due to the simplicity of this transition matrix, it is possible to calculate all of the terms Eqs.~\eqref{eq:expansion_terms1}-\eqref{eq:expansion_terms} analytically and subsequently integrate Eq.~\eqref{eq:stotint} to get the optimal protocols for different $t_f$'s. This calculation can be done exactly, but the equations are very lengthy and are presented in appendix \ref{app:derivative_W_dagger} and \ref{app:rewrite_W_dagger}, and in the Mathematica notebook available at \footnote{\texttt{https://github.com/KarelProesmans/Nonequilibrium\\\_control}}

\begin{figure}
	\centering
    \includegraphics[width=0.9\linewidth]{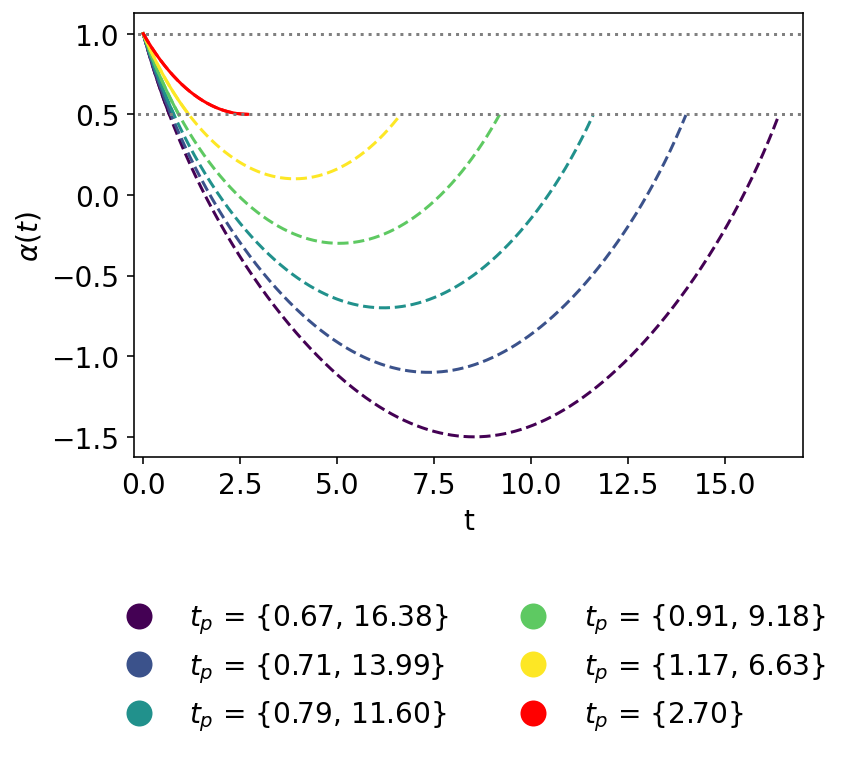}
	
	\caption{\textbf{Examples of optimal protocols of various duration, $\boldsymbol{t_p}$.} The grey, dotted lines indicate initial and final values of $\alpha$. The short, solid protocols in the upper left corner are the 'direct protocols', where the system moves monotonously from $\alpha_i$ to $\alpha_f$. The broken curves mark the 'overshoot protocols'. These protocols are continuations of the direct protocols. After passing $\alpha_f$ the first time, $\alpha$ continues to decrease, then switches and returns back to $\alpha_f$. This added detour turns out to be entropically favorable at longer protocol durations.}
	\label{fig:optimal_protocols}
\end{figure}
We will immediately go to the results. Firstly, a subset of the optimal protocols, found by integrating Eq.~\eqref{eq:full_df_eq}, is shown in fig.~\ref{fig:optimal_protocols}. We see that there are two types of protocols: for protocols of durations shorter than
\begin{equation}
    t^*_p=\int^{\alpha_f}_{\alpha_0}d\alpha\left(\sqrt{\frac{A^{(1)}(\alpha)+A^{(2)}(\alpha)-A^{(3)}(\alpha)}{\sigma^{SS}(\alpha)-\sigma^{SS}_{\textrm{min}}}}\right)\approx 2.70
\end{equation}
where $\sigma^{SS}_{\textrm{min}}=\sigma^{SS}(1/2)$, the minimum of $\sigma^{SS}$ for $\alpha$ between $1/2$ and $1$, and protocols longer than $t^*_p$. For $t_p<t^*_p$, the protocols move $\alpha$ monotonously from $\alpha=1$ to $\alpha=0.5$, whereas for $t_p>t_p^*$, the protocol initially follows the same route as a shorter protocol, but then undershoots to eventually return to the final value $\alpha=0.5$.

\begin{figure}
    \centering
    \includegraphics[width=0.9\linewidth]{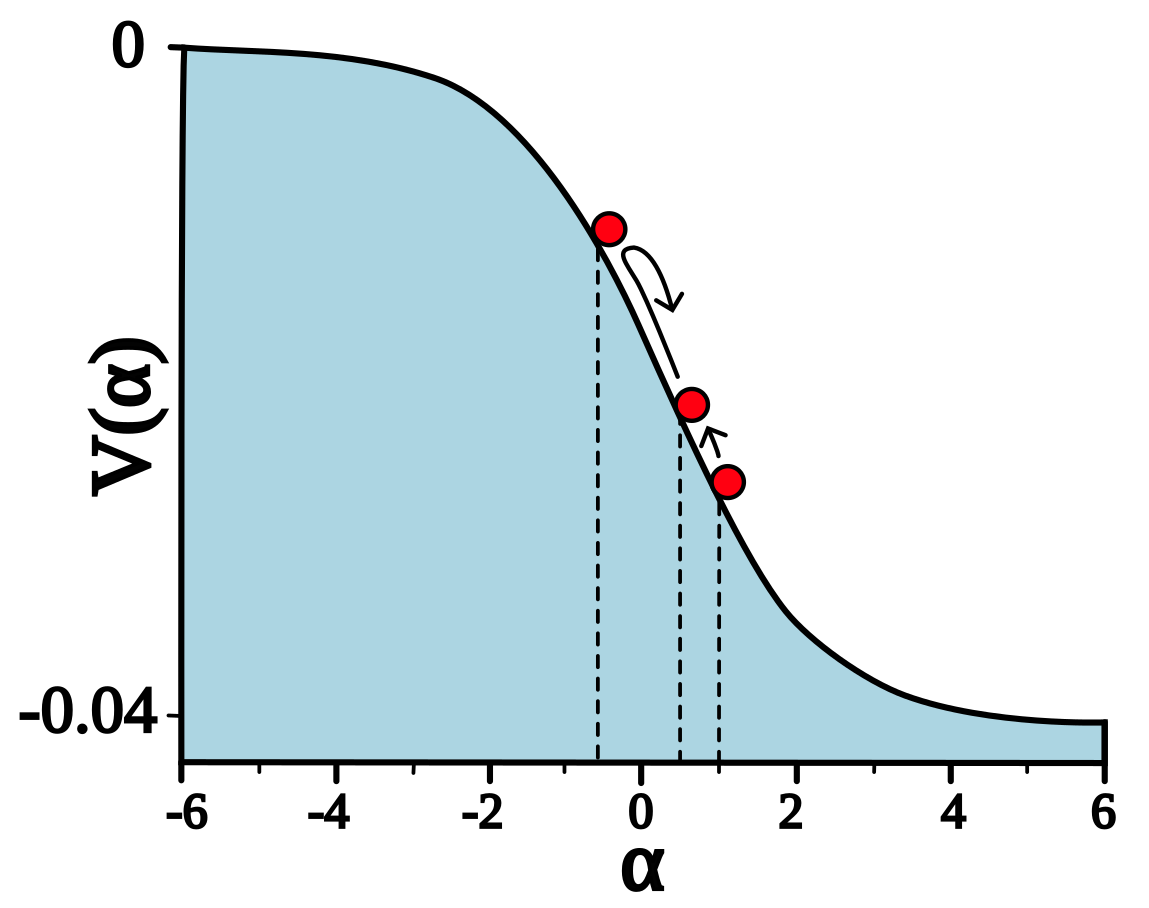}
    \includegraphics[width=0.9\linewidth]{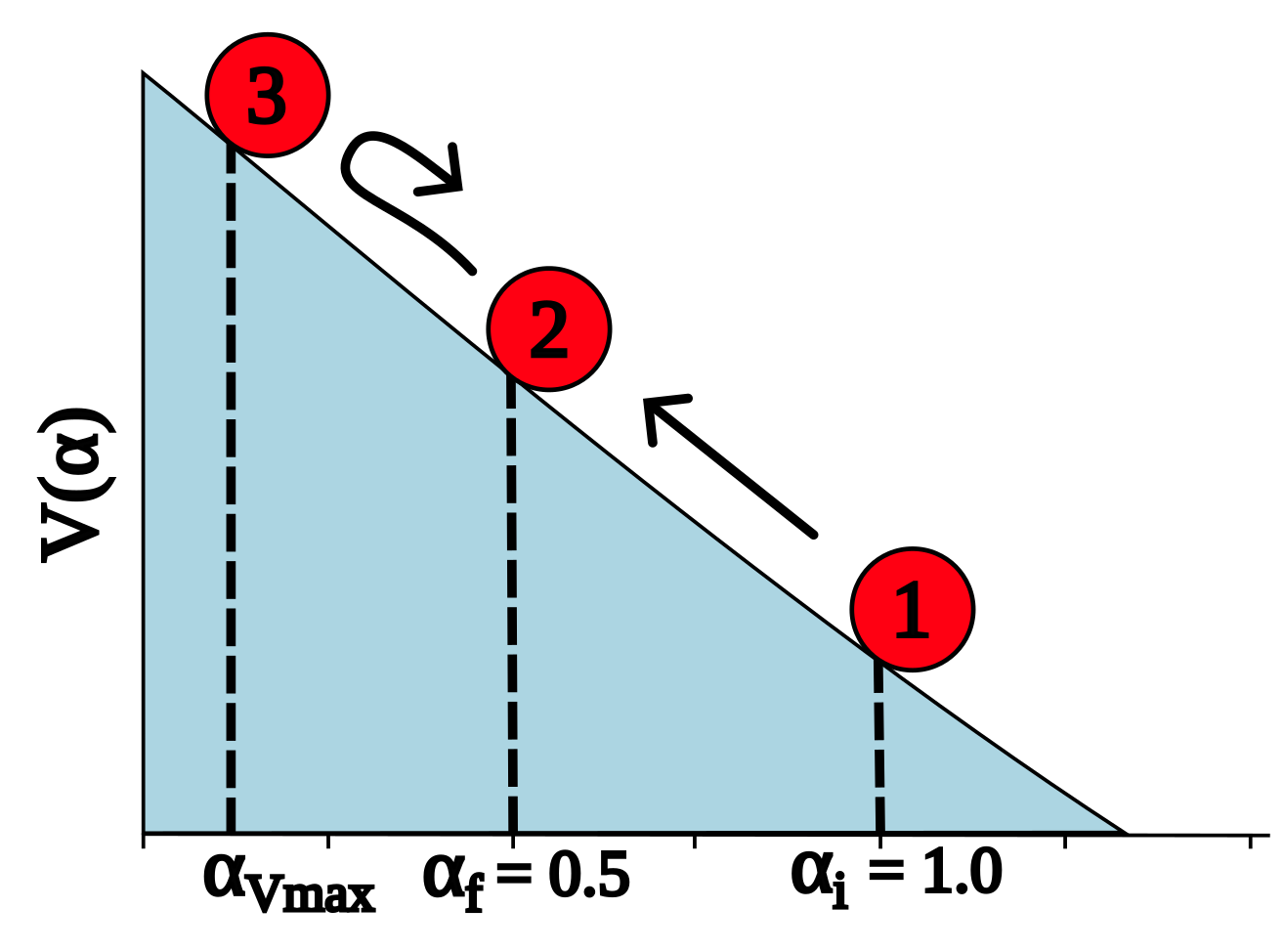}
    \caption{\textbf{The potential landscape.} The top drawing depicts the full shape of the potential landscape, given as $V(\alpha) = -\sigma_{SS}(\alpha)$. As the system moves uphill, the steady state dissipation decreases. The 'point mass' or red dot is the three state system. The bottom drawing is a close-up of the slope. During direct protocols, the system travels from 1 to 2, while overshoot protocols travel $1 \rightarrow 2 \rightarrow 3 \rightarrow 2$. Here we see how a large overshoot implies spending most of the protocol far uphill at low steady state entropy production. Since steady state is the dominating contribution to the overall entropy production in this range of $t_p$, this is an optimal strategy (see appendix \ref{app:terms_plots}). }
    \label{fig:potential}
\end{figure}

This second type of protocols has not been observed in similar frameworks of systems near equilibrium \cite{sivak_thermodynamic_2012} or when only taking into account part of the entropy production \cite{mandal_analysis_2016}. To understand why this happens, we can think of the Lagrangian, Eq.~\eqref{eq:Lres} as that of a mechanical systems, such as a particle at position $\alpha$ in a potential energy landscape given by $V(\alpha)=-\sigma_{SS}(\alpha)$ and positional dependent mass and friction coefficient, cf.~Fig.~\ref{fig:potential}. The optimal protocol then corresponds to the dynamical trajectory of this mechanical system. In the near-equilibrium picture, one would have $V(\alpha)=0$, due to the absence of steady-state entropy production. Meanwhile, in the mechanical picture, it is clear that the system can overshoot and then return to the original value if $\sigma_{SS}(\alpha)$ is non-zero.

\begin{figure}
	\centering
	\includegraphics[width=0.9\linewidth]{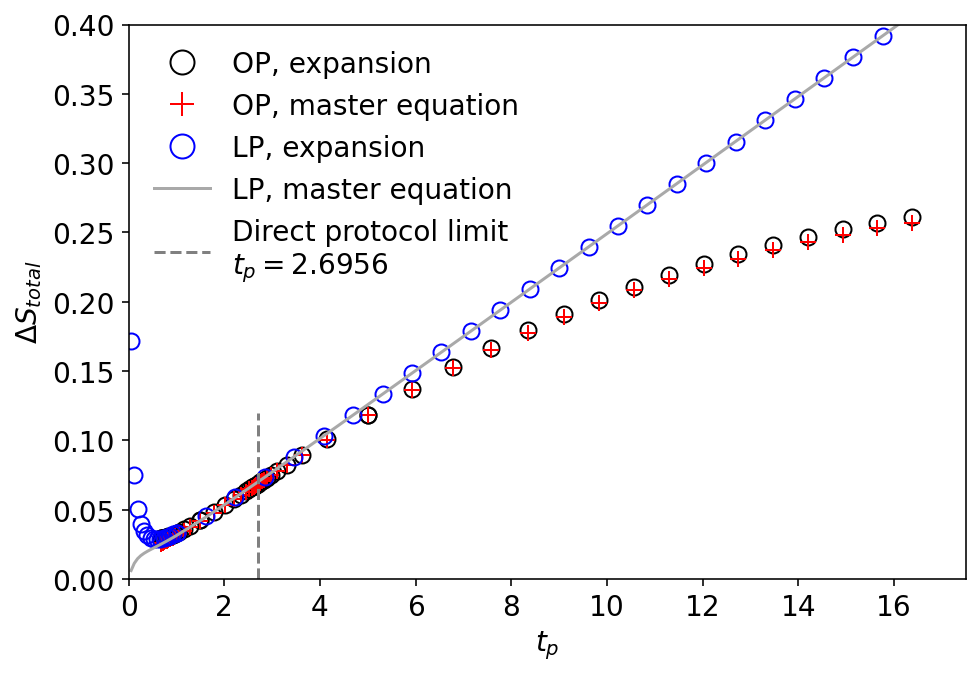}
	\caption{\textbf{Total entropy of optimized protocols, various durations, $\boldsymbol{t_p}$.} Optimal protocols are marked 'OP', linear protocols are marked 'LP'. The entropy obtained by integrating the expansion is compared to the entropy obtained by direct Euler integration of the master equation. The optimal 'overshoot' protocol produces less entropy than the linear protocol at long protocol durations. Interestingly, the expansion and numerical integration of the optimal protocol deviates at longer protocol duration. This feature is related to the rate of the protocol $\dot \alpha$ not being sufficiently low (see appendix \ref{app:expansion_deviation}).}
	\label{fig:all_dissipation}
\end{figure}

One might wonder at this point how well these protocols are at minimising the entropy production. To answer this question, we compare the entropy production of the optimal protocols with those of linear protocols,
\begin{eqnarray}	
	\alpha(t) = \frac{\alpha_f-\alpha_i}{t_p} \cdot t + \alpha_i, \quad \mathrm{and} \quad t_f - t_i = t_p
\end{eqnarray}
The results are shown in fig.~\ref{fig:all_dissipation}.

We can draw several conclusions from this. Firstly, we do see that the optimal protocol does indeed always produce less entropy than the linear protocol. The difference seems small for short durations, but becomes more pronounced for longer protocols. In particular, we see that the entropy production of the linear protocol grows linearly with protocol duration, whereas the entropy production of the optimal protocol levels off for long protocol durations. Secondly, we see that the entropy porduction calculated from Eq.~\eqref{eq:stotint} and the numerically exact entropy production up to rather short protocol duration ($t_p\gtrsim 1$).\\



\section{\label{sec:discussion} Discussion}

In conclusion, we have developed a framework to minimise the entropy production of slowly driven systems far from thermodynamic equilibrium. We have found that both the optimal protocols and the minimal entropy production are fundamentally different from systems that operate close to equilibrium.

For future research, it would be interesting to test our results experimentally, similar to what has been done for near-equilibrium systems \cite{tafoya_using_2019}. Furthermore, our framework can be used to study the thermodynamics of non-equilibrium phase-transitions \cite{herpich2018collective,nguyen_exponential_2020,pham2024irreversibility}. In particular, it would be interesting to see how one can minimise entropy production, while driving a system through such a phase-transition.

\appendix
    
    \section{\label{app:expand_state_vec} Expanding $\mathbf{p}(t)$} 
    Following \cite{mandal_analysis_2016}, we derive the expansion for the state probability vector by rephrasing the master equation in terms of $\mathbf{p}_{\mathrm{SS}} (t)$ and $\Delta \mathbf{p}(t)=\mathbf{p}(t)-\mathbf{p}_{\mathrm{SS}} (t)$.\\

First, we differentiate the above expression and use the master equation (See eq.~\ref{eq:master_eq}): 

\begin{eqnarray}
	\frac{d\mathbf{p}_{\mathrm{SS}}(t) }{dt} &= \frac{d}{dt} \left( \mathbf{p}(t)  - \Delta \mathbf{p}(t) \right) \nonumber\\
	&= W \mathbf{p}(t)  - \frac{d}{d t} \Delta \mathbf{p}(t)
\end{eqnarray}

We now add $W \mathbf{p}_\mathrm{SS} = 0$ (Definition of steady state) and exchange $\mathbf{p}(t)$ with $ \Delta \mathbf{p}(t)$:

\begin{eqnarray}
	\frac{d\mathbf{p}_{\mathrm{SS}}(t) }{dt}&=& W (\mathbf{p}(t) -\mathbf{p}_{\mathrm{SS}}(t) )   - \frac{d}{d t} \Delta \mathbf{p}(t) \nonumber\\
	&=& -\frac{d}{d t} \Delta \mathbf{p}(t)  + W  \Delta \mathbf{p}(t) 
\end{eqnarray}

Using the pseudo-inverse $W^\dagger$ on both sides, the equation is simplified:
\begin{eqnarray}
	  W^\dagger \frac{d \mathbf{p}_{\mathrm{SS}}(t) }{dt} &=& -W^\dagger \frac{d}{dt} \Delta \mathbf{p}(t)  + W^\dagger W  \Delta \mathbf{p}(t) \nonumber\\
	&=& -W^\dagger \frac{d}{dt} \Delta \mathbf{p}(t) + \left( 1 - \vert \mathbf{p}_{\mathrm{SS}} \rangle \langle 1 \vert \right)  \Delta \mathbf{p}(t) \nonumber \\
    &=&\left[ 1 -  W^\dagger \frac{d}{dt} \right] \Delta \mathbf{p}(t)
\end{eqnarray}

Part of the second term on the left was eliminated, since: 

\begin{eqnarray}
	\vert \mathbf{p}_{\mathrm{SS}} \rangle \langle 1 \vert \Delta \mathbf{p}(t) = \vert \mathbf{p}_{\mathrm{SS}} \rangle \left( \langle 1 \vert \mathbf{p}(t) - \langle 1 \vert  \mathbf{p}_{\mathrm{SS}}(t) \right)= 0\nonumber\\
\end{eqnarray}

This equation has the following solution for $\Delta \mathbf{p}$ \cite{mandal_analysis_2016}: 

\begin{eqnarray}
	\Delta \mathbf{p} = \sum_{n=1}^{\infty}\left( W^\dagger  \frac{d}{dt} \right)^n \mathbf{p}_{\mathrm{SS}}
\end{eqnarray}
    ´
    \section{\label{app:derivative_W_dagger} Partial derivative of $W^\dagger$} 
    The partial derivative of $W^\dagger$, $\partial W/\partial \alpha $, is found by partially differentiating the expression $\partial(WW^\dagger)/\partial \alpha $:

\begin{eqnarray}
	W\frac{\partial W^\dagger}{\partial \alpha} = \frac{\partial }{\partial \alpha} (WW^\dagger) - \left(\frac{\partial W}{\partial \alpha}\right) W^\dagger
\end{eqnarray}

Multiplying on both sides with the pseudo-inverse of $W$, we have:

\begin{eqnarray}
	W^\dagger W &&\frac{\partial W^\dagger}{\partial \alpha} =\nonumber\\ &&W^\dagger\frac{\partial}{\partial \alpha} (\mathbf{1}  - \vert \mathbf{p}_{\mathrm{SS}} \rangle \langle 1 \vert) - W^\dagger \left(\frac{\partial W}{\partial \alpha}\right) W^\dagger
\end{eqnarray}

From the definition of $W^\dagger$, $W^\dagger W$ is replaced by the identity matrix and the outer product of the 0-value eigenvectors. Since $W^\dagger$ share the same 0th eigenvectors as $W$, $ \langle 1 \vert$ and $\vert \mathbf{p}_{\mathrm{SS}} \rangle$, the second term on the left disappears:

\begin{eqnarray}
	(\mathbf{1}  -\vert \mathbf{p}_{\mathrm{SS}} \rangle \langle 1 \vert) \frac{\partial W^\dagger}{\partial \alpha}=\qquad&&\\ 
    W^\dagger (0 - \frac{\partial}{\partial \alpha}&& \vert \mathbf{p}_{\mathrm{SS}} \rangle \langle 1 \vert) - W^\dagger \left(\frac{\partial W}{\partial \alpha}\right) W^\dagger \nonumber\\
	\Leftrightarrow \frac{\partial W^\dagger}{\partial \alpha} - \vert \mathbf{p}_{\mathrm{SS}} \rangle \frac{\partial \langle 1 \vert W^\dagger}{\partial \alpha} &&= \nonumber\\
    - W^\dagger \frac{\partial}{\partial \alpha} \vert \mathbf{p}_{\mathrm{SS}} \rangle&& \langle 1 \vert - W^\dagger \left(\frac{\partial W}{\partial \alpha}\right) W^\dagger \\
	\Leftrightarrow \frac{\partial W^\dagger}{\partial \alpha} = - W^\dagger \frac{\partial}{\partial \alpha} \vert \mathbf{p}_{\mathrm{SS}}&& \rangle \langle 1 \vert - W^\dagger \left(\frac{\partial W}{\partial \alpha}\right) W^\dagger
\end{eqnarray}

    \section{\label{app:expansion_terms} Expansion terms} 






The first step is to express the state probability vector $\mathbf{p}(t)$ in terms of $\mathbf{p}_{\mathrm{SS}}(t)$ using the expansion in \cite{mandal_analysis_2016} (see appendix \ref{app:expand_state_vec}). For simplicity, we start out with the case of a single control parameter, $\alpha(t)$.

\begin{eqnarray}
	\mathbf{p}(t) &=& \sum_{0}^{\infty}\left(W^\dagger\frac{d}{dt}\right)^n\mathbf{p}_{\mathrm{SS}}(t) \nonumber\\
	&=&\sum_{0}^{\infty}\left(W^\dagger\left[\frac{\partial}{\partial t} + \dot\alpha\frac{\partial}{\partial \alpha}\right]\right)^n\mathbf{p}_{\mathrm{SS}}(t)
    \label{eq:appendix_p_expansion_inf}
\end{eqnarray}

The zeroth order of this expansion is given by:

\begin{eqnarray}
	\mathbf{p}^{(0)} = \mathbf{p}_{\mathrm{SS}}(t)
\end{eqnarray}

$\mathbf{p}(t)$ to 1st order is: 

\begin{eqnarray}
	W^\dagger&&\left[\frac{\partial}{\partial t} + \dot\alpha\frac{\partial}{\partial \alpha}\right]\mathbf{p}_{\mathrm{SS}}(t) \nonumber\\
    &&= W^\dagger\dot\alpha\frac{\partial \mathbf{p}_{\mathrm{SS}}(t)}{\partial \alpha} = \dot\alpha \mathbf{p}^{(1)}
\end{eqnarray}

Here we note that $\mathbf{p}_{\mathrm{SS}}(t)$ is not explicitly dependent on $t$, but only through $\alpha(t)$.

The 2nd order is a little more elaborate:
\begin{eqnarray}
    W^\dagger&&\left[\frac{\partial}{\partial t} + \dot\alpha\frac{\partial}{\partial \alpha}\right]W^\dagger\left[\frac{\partial}{\partial t} + \dot\alpha\frac{\partial}{\partial \alpha}\right]\mathbf{p}_{\mathrm{SS}}(t) = \\
	&&W^\dagger\frac{\partial}{\partial t} \left( W^\dagger\frac{\partial \mathbf{p}_{\mathrm{SS}}(t)}{\partial t} \right) 
	+ W^\dagger\frac{\partial}{\partial t} \left(W^\dagger \dot\alpha\frac{\partial \mathbf{p}_{\mathrm{SS}}(t)}{\partial \alpha} \right)\nonumber\\
	&&+ W^\dagger\dot\alpha\frac{\partial}{\partial \alpha}\left( W^\dagger \frac{\partial \mathbf{p}_{\mathrm{SS}}(t)}{\partial t} \right) 
	+ W^\dagger\dot\alpha\frac{\partial}{\partial \alpha}\left(W^\dagger \dot\alpha\frac{\partial \mathbf{p}_{\mathrm{SS}}(t)}{\partial \alpha}\right)\nonumber
\end{eqnarray}
    
Many terms disappear due to the lack of explicit time dependence of both rates, $W$, and the steady-state distribution:

\begin{eqnarray}
    \left(W^\dagger\left[\frac{\partial}{\partial t} + \dot\alpha\frac{\partial}{\partial \alpha}\right] \right)^2\mathbf{p}_{\mathrm{SS}}(t) =& \\
	W^\dagger\frac{\partial}{\partial t} \left(W^\dagger \dot\alpha\frac{\partial \mathbf{p}_{\mathrm{SS}}(t)}{\partial \alpha} \right)&
	+ W^\dagger\dot\alpha\frac{\partial}{\partial \alpha}\left(W^\dagger \dot\alpha\frac{\partial \mathbf{p}_{\mathrm{SS}}(t)}{\partial \alpha}\right)\nonumber
\end{eqnarray}
    
We continue expanding and removing terms involving explicit time-dependency:
    \begin{eqnarray}
    \left(W^\dagger\left[\frac{\partial}{\partial t} + \dot\alpha\frac{\partial}{\partial \alpha}\right] \right)^2\mathbf{p}_{\mathrm{SS}}(t) &&= \nonumber\\
	W^\dagger\left(\frac{\partial W^\dagger}{\partial t}  \dot\alpha\frac{\partial \mathbf{p}_{\mathrm{SS}}(t)}{\partial \alpha} \right)&& + W^\dagger\left(W^\dagger\ddot\alpha\frac{\partial \mathbf{p}_{\mathrm{SS}}(t)}{\partial \alpha} \right) \nonumber\\
	\quad + W^\dagger\left(W^\dagger\dot\alpha\frac{\partial^2\mathbf{p}_{\mathrm{SS}}(t)}{\partial \alpha\partial t} \right)&&
	+ W^\dagger\dot\alpha\frac{\partial}{\partial \alpha}\left(W^\dagger \dot\alpha\frac{\partial \mathbf{p}_{\mathrm{SS}}(t)}{\partial \alpha}\right)\nonumber\\
	=\left(W^\dagger\right)^2\ddot\alpha\frac{\partial \mathbf{p}_{\mathrm{SS}}(t)}{\partial \alpha}
	+ W^\dagger&&(\dot\alpha)^2\frac{\partial}{\partial \alpha}\left(W^\dagger \frac{\partial \mathbf{p}_{\mathrm{SS}}(t)}{\partial \alpha}\right)\\
	=\left(W^\dagger\right)^2\ddot\alpha\frac{\partial \mathbf{p}_{\mathrm{SS}}(t)}{\partial \alpha}
	+ W^\dagger&&(\dot\alpha)^2\left(\frac{\partial W^\dagger}{\partial \alpha} \frac{\partial \mathbf{p}_{\mathrm{SS}}(t)}{\partial \alpha} + W^\dagger\frac{\partial^2 \mathbf{p}_{\mathrm{SS}}(t)}{\partial \alpha^2}\right) \nonumber
\end{eqnarray}

Put together, the 2nd-order expansion of the state-probability vector is given by:

\begin{eqnarray}
	\mathbf{p}(t) \approx 
	\mathbf{p}_{\mathrm{SS}}(t)&& + \dot\alpha \cdot W^\dagger \frac{\partial}{\partial \alpha}\mathbf{p}_{\mathrm{SS}}(t)\nonumber\\
    &&+ (\dot\alpha)^2  W^\dagger\left(\frac{\partial W^\dagger}{\partial \alpha}\frac{\partial \mathbf{p}_{\mathrm{SS}}(t)}{\partial \alpha} 
    + W^\dagger\frac{\partial^2 \mathbf{p}_{\mathrm{SS}}(t)}{\partial \alpha^2}\right) \nonumber\\
    &&+ \ddot\alpha \cdot \left(W^\dagger\right)^2\frac{\partial \mathbf{p}_{\mathrm{SS}}(t)}{\partial \alpha}
      \label{eq:appendix_p_expansion_2nd}
\end{eqnarray}

On the form:

\begin{eqnarray}
	\mathbf{p}(t) \approx 
	\mathbf{p}^{(0)}\
	+ \dot\alpha \cdot \mathbf{p}^{(1)} 
	+ (\dot\alpha)^2 \cdot \mathbf{p}^{(2)} 
	+ \ddot\alpha \cdot \mathbf{p}^{(3)}
\end{eqnarray}

Where the last two terms are both contributions of second order in $\epsilon$, and can be
grouped together:

\begin{eqnarray}
	\mathbf{p}(t) \approx 
	p^{(0)}
	+ \epsilon \cdot \mathbf{p}^{(1)} 
	+ \epsilon^2 \cdot \left( \mathbf{p}^{(2)} + \mathbf{p}^{(3)} \right) \quad \epsilon \propto \dot\alpha \nonumber \\
\end{eqnarray}

We then proceed to expand $\sigma$ (see eq.~\ref{eq:entr_prod}) as a function of $\epsilon \propto \dot\alpha$ around $\mathbf{p}(t) =\mathbf{p}_{\mathrm{SS}}(t)$. The derivatives that appear, $\partial \mathbf{p} /\partial \epsilon$ and $\partial^2 \mathbf{p} /\partial \epsilon^2$, can then be related to $\mathbf{p}(t)$ through the steady-state expansion (See eq.~\ref{eq:appendix_p_expansion_2nd}). For simplicity, we use the short hand notation $\{\mathbf{p}^{(0)}, \mathbf{p}^{(1)}, \mathbf{p}^{(2)}\}$.

\begin{eqnarray}
	\sigma = k_B \sum_{i,j,(\nu)} W_{ij}^{(\nu)}p_j\ln\frac{W_{ij}^{(\nu)}p_j}{W_{ji}^{(\nu)}p_i}
\end{eqnarray}

\begin{eqnarray}
	\sigma = \sigma \vert_{\mathbf{p} = \mathbf{p}_{\mathrm{SS}}} + \epsilon \frac{\partial \sigma}{\partial \epsilon}\vert_{\mathbf{p} = \mathbf{p}_{\mathrm{SS}}} + \epsilon^2 \frac{1}{2}\frac{\partial^2 \sigma}{\partial \epsilon^2}\vert_{\mathbf{p} = \mathbf{p}_{\mathrm{SS}}} + \mathcal{O}(\epsilon^3)\nonumber \\
\end{eqnarray}

The 0th term is easily evaluated:

\begin{eqnarray}
	\sigma \vert_{\mathbf{p} = \mathbf{p}_{\mathrm{SS}}}  = k_B \sum_{i,j,(\nu)} W_{ij}^{(\nu)}p_{\mathrm{SS}, j}\ln\frac{W_{ij}^{(\nu)}p_{\mathrm{SS}, j}}{W_{ji}^{(\nu)}p_{\mathrm{SS}, i}}
\end{eqnarray}

The 1st term is a little more elaborate:

\begin{eqnarray}
    \frac{\partial \sigma}{\partial \epsilon}&&\vert_{\mathbf{p} = \mathbf{p}_{\mathrm{SS}}} = k_B \sum_{i, j, (\nu)} W_{ij}^{(\nu)} \nonumber\\
    && \quad \times \left( \frac{\partial p_j}{\partial \epsilon}\ln\frac{W_{ij}^{(\nu)}p_j}{W_{ji}^{(\nu)}p_i} + p_j \left( p_j ^{-1} \frac{\partial p_j}{\partial \epsilon} - p_i ^{-1} \frac{\partial p_i}{\partial \epsilon}\right)\right)\vert_{\mathbf{p} = \mathbf{p}_{\mathrm{SS}}} \nonumber \\
    &&= k_B \sum_{i, j, (\nu)} W_{ij}^{(\nu)} \nonumber\\
    &&  \quad  \times\left(p_j^{(1)}\ln\frac{W_{ij}^{(\nu)}p_{\mathrm{SS}, j}}{W_{ji}^{(\nu)}p_{\mathrm{SS}, i}} + p_{\mathrm{SS}, j} \left( p_{\mathrm{SS}, j} ^{-1} p_j^{(1)} - p_{\mathrm{SS}, i} ^{-1} p_i^{(1)} \right)\right) \nonumber\\
    &&= k_B \sum_{i, j, (\nu)} W_{ij}^{(\nu)}p_j^{(1)}\ln\frac{W_{ij}^{(\nu)}p_{\mathrm{SS}, j}}{W_{ji}^{(\nu)}p_{\mathrm{SS}, i}}
\end{eqnarray}

The latter terms cancel out when summing, since the properties of $\langle 1 \,\vline$ and $\vline \, p_{\mathrm{SS}} \rangle$ ensures that:

\begin{itemize}
    \item[(1)] $\sum_i W_{ij} = 0$
    \item[(2)] $\sum_j W_{ij} p_{\mathrm{SS}, j} = 0$
\end{itemize} 

\begin{widetext}
Finally, for the 2nd term, first twice differentiating $\sigma$:
\begin{eqnarray}
    \frac{\partial^2 \sigma}{\partial \epsilon^2} = k_B \sum_{i,j,(\nu)}\ W_{ij}^{(\nu)}\Bigg[\frac{\partial ^2 p_j}{\partial \epsilon^2}\ln\frac{W_{ij}^{(\nu)}p_j}{W_{ji}^{(\nu)}p_i}
    &&+2\frac{\partial p_j}{\partial \epsilon}\left( p_j ^{-1}\frac{\partial p_j}{\partial \epsilon} - p_i^{-1}\frac{\partial p_i}{\partial \epsilon}\right)\nonumber\\
    &&+ p_j\left( p_j^{-1}\frac{\partial^2 p_j}{\partial \epsilon^2} -p_j^{-2}\left(\frac{\partial  p_j}{\partial \epsilon}\right)^2- 
    p_i^{-1}\frac{\partial^2 p_i}{\partial \epsilon^2} + p_i^{-2}\left(\frac{\partial p_i}{\partial \epsilon}\right)^2\right)\Bigg]
\end{eqnarray}

And then evaluating at steady state:
 
\begin{eqnarray}
	\frac{\partial^2 \sigma}{\partial \epsilon^2}\vert_{\mathbf{p} = \mathbf{p}_{\mathrm{SS}}} = k_B&& \sum_{i,j,(\nu)}\ W_{ij}^{(\nu)}\Bigg[2\left(p_j^{(2)} + p_j^{(3)} \right)\ln\frac{W_{ij}^{(\nu)}p_{\mathrm{SS}, j}}{W_{ji}^{(\nu)}p_{\mathrm{SS}, i}} + 2p_j^{(1)}\left( p_{\mathrm{SS}, j} ^{-1}p_j^{(1)} - p_{\mathrm{SS}, i}^{-1}p_i^{(1)}\right) \nonumber\\
    &&+ p_{\mathrm{SS}, j}\left( 2p_{\mathrm{SS}, j}^{-1} \left(p_j^{(2)} + p_j^{(3)} \right)- p_{\mathrm{SS}, j}^{-2}\left(p_j^{(1)}\right)^2 - 2p_{\mathrm{SS}, i}^{-1} \left(p_i^{(2)} + p_i^{(3)} \right) + p_{\mathrm{SS}, i}^{-2}\left(p_i^{(1)}\right)^2\right)\Bigg] \nonumber\\
	&&= k_B \sum_{i,j,(\nu)}\ W_{ij}^{(\nu)}\Bigg[2 \left(p_j^{(2)} + p_j^{(3)} \right)\ln\frac{W_{ij}^{(\nu)}p_{\mathrm{SS}, j}}{W_{ji}^{(\nu)}p_{\mathrm{SS}, i}}\
	-2p_i^{(1)} p_{\mathrm{SS}, j}^{-1}p_j^{(1)}\Bigg]
\end{eqnarray}

Note the change in index in the last term in the square brackets. Similarly to before, many terms cancel due to the properties of $\langle 1 \, \vline$ and $\vline  \,p_{\mathrm{SS}} \rangle$.\\

Gathering all the terms, we have:

\begin{eqnarray}
	\sigma = 
	k_B&&\sum_{i,j,\nu} W_{ij}^{(\nu)}p_{\mathrm{SS}, j}\ln\frac{W_{ij}^{(\nu)}p_{\mathrm{SS}, j}}{W_{ji}^{(\nu)}p_{\mathrm{SS}, i}} + \epsilon \cdot k_B \sum_{i, j, (\nu)} W_{ij}^{(\nu)}p_j^{(1)}\ln\frac{W_{ij}^{(\nu)}p_{\mathrm{SS}, j}}{W_{ji}^{(\nu)}p_{\mathrm{SS}, i}} \nonumber\\
	&&+ \epsilon^2 \cdot \frac{1}{2} k_B \sum_{i,j,(\nu)}W_{ij}^{(\nu)}\Bigg[2\left(p_j^{(2)} + p_j^{(3)} \right)\ln\frac{W_{ij}^{(\nu)}p_{\mathrm{SS}, j}}{W_{ji}^{(\nu)}p_{\mathrm{SS}, i}}
	-2p_i^{(1)} p_{\mathrm{SS}, j}^{-1}p_j^{(1)}\Bigg] + \mathcal{O}(\epsilon^3)
\end{eqnarray}

We then substitute $\mathbf{p}^{(0)}, \mathbf{p}^{(1)}, \mathbf{p}^{(2)}$ and $\mathbf{p}^{(3)}$ for the corresponding terms in the Mandal \& Jarzynski expansion of $\mathbf{p}(t)$ (See eq.~\ref{eq:appendix_p_expansion_2nd}): 

\begin{eqnarray}
	\sigma = k_B \sum_{i,j,(\nu)} W_{ij}^{(\nu)}p_{\mathrm{SS}, j}\ln\frac{W_{ij}^{(\nu)}p_{\mathrm{SS}, j}}{W_{ji}^{(\nu)}p_{\mathrm{SS}, i}}
	+ \epsilon \cdot k_B \sum_{i, j, k, (\nu)} &&\ln\frac{W_{ij}^{(\nu)}p_{\mathrm{SS}, j}}{W_{ji}^{(\nu)}p_{\mathrm{SS}, i}}W_{ij}^{(\nu)}W_{jk}^\dagger\frac{\partial p_{\mathrm{SS}, k}}{\partial \alpha} \nonumber\\
	+ \epsilon^2 \cdot \frac{1}{2} k_B \sum_{i,j, k, l, (\nu)}W_{ij}^{(\nu)}\
	\Bigg[2\ln\frac{W_{ij}^{(\nu)}p_{\mathrm{SS}, j}}{W_{ji}^{(\nu)}p_{\mathrm{SS}, i}}
    \cdot &&\left( W_{jk}^\dagger\left(\frac{\partial W_{kl}^\dagger}{\partial \alpha}\frac{\partial p_{\mathrm{SS}, l}}{\partial \alpha} 
	+ W_{kl}^\dagger\frac{\partial^2 p_{\mathrm{SS}, l}}{\partial \alpha^2}\right) +  W^\dagger_{jk}W^\dagger_{kl}\frac{\partial p_{\mathrm{SS}, l}}{\partial \alpha} \right) \nonumber\\
	&&-2 W_{ik}^\dagger \frac{\partial p_{\mathrm{SS}, k}}{\partial \alpha}  p_{\mathrm{SS}, j}^{-1} W^\dagger_{jl} \frac{\partial p_{\mathrm{SS}, l}}{\partial \alpha}\Bigg] + \mathcal{O}(\epsilon^3)
\end{eqnarray}

Simplifying further:



\begin{eqnarray}
	\sigma = 
	k_B \sum_{i,j,(\nu)}&& \ln\frac{W_{ij}^{(\nu)}p_{\mathrm{SS}, j}}{W_{ji}^{(\nu)}p_{\mathrm{SS}, i}}W_{ij}^{(\nu)}p_{\mathrm{SS}, j}
	+ \epsilon \cdot k_B \sum_{i, j, k, (\nu)} \ln\frac{W_{ij}^{(\nu)}p_{\mathrm{SS}, j}}{W_{ji}^{(\nu)}p_{\mathrm{SS}, i}}W_{ij}^{(\nu)}W_{jk}^\dagger\frac{\partial p_{\mathrm{SS}, k}}{\partial \alpha}\nonumber\\
	&&+ \epsilon^2 \cdot k_B \sum_{i,j, k, l, (\nu)} \ln\frac{W_{ij}^{(\nu)}p_{\mathrm{SS}, j}}{W_{ji}^{(\nu)}p_{\mathrm{SS}, i}} W_{ij}^{(\nu)}\ W_{jk}^\dagger\frac{\partial}{\partial \alpha}\left( W_{kl}^\dagger \frac{\partial p_{\mathrm{SS}, l}}{\partial \alpha}\right) + \epsilon^2 \cdot k_B \sum_{i,j, k, l, (\nu)} \ln\frac{W_{ij}^{(\nu)}p_{\mathrm{SS}, j}}{W_{ji}^{(\nu)}p_{\mathrm{SS}, i}} W_{ij}^{(\nu)}\ W^\dagger_{jk}W^\dagger_{kl}\frac{\partial p_{\mathrm{SS}, l}}{\partial \alpha} \nonumber\\
	&&-\epsilon^2 \cdot k_B \sum_{i,j,(\nu)}\frac{\partial p_{\mathrm{SS}, i}}{\partial \alpha} W_{ij}^\dagger p_{\mathrm{SS}, j}^{-1} \frac{\partial p_{\mathrm{SS}, j}}{\partial \alpha} + \mathcal{O}(\epsilon^3)
\end{eqnarray}

Note that $\sum_{j} W_{ij} W^\dagger_{jl} = \delta_{il} - p_{\mathrm{SS}, i} $ and $\sum_l  \frac{\partial p_{\mathrm{SS}, l}}{\partial \alpha}  = 0$. This leads to the results from the main text.

\end{widetext}

This framework can be expanded to several parameters: $\boldsymbol{\alpha}(t) = \{\alpha_{a_{1}}(t), \alpha_{a_{2}}(t), .\,.\,.\,\}$.

When expanding $\mathbf{p}(t)$, the time derivative is given by:
$d/dt = \partial/\partial t +  \boldsymbol{\dot\alpha} \cdot \nabla_{\boldsymbol{\alpha}}$:

\begin{eqnarray}
	\mathbf{p}(t) =\sum_{0}^{\infty}\left(W^\dagger\left[\frac{\partial}{\partial t} + \sum_a\dot\alpha_a\frac{\partial}{\partial \alpha_a}\right]\right)^n\mathbf{p}_{\mathrm{SS}}(t) \quad
    \label{eq:appendix_p_expansion_inf_mult_par}
\end{eqnarray}

$\mathbf{p}(t)$ to second order follows a similar derivation as for one parameter:

\begin{eqnarray} 
    \mathbf{p}(t) \approx \mathbf{p}^{(0)}&& + \sum_a \dot\alpha_a \cdot \mathbf{p}^{(1)}_a \nonumber\\
    &&+ \sum_{a,b} \dot\alpha_a \dot\alpha_b \mathbf{p}^{(2)}_{ab} + \sum_a \ddot \alpha_a \mathbf{p}^{(3)}_a
\end{eqnarray}

on the form:

\begin{eqnarray} 
    \mathbf{p}(t) \approx \mathbf{p}^{(0)}&& + \sum_a \epsilon_a \cdot \mathbf{p}^{(1)}_a \\
    &&+ \sum_{a,b} \epsilon_a \epsilon_b \mathbf{p}^{(2)}_{ab} + \sum_a \epsilon_a^2 \mathbf{p}^{(3)}_a
\end{eqnarray}

0th order:

\begin{eqnarray}
    \mathbf{p}^{(0)} = \mathbf{p}_{\mathrm{SS}}(t)
\end{eqnarray}

1st order:

\begin{eqnarray}
    W^\dagger&&\left[\frac{\partial}{\partial t} + \sum_a\dot\alpha_a\frac{\partial}{\partial \alpha_a}\right]\mathbf{p}_{\mathrm{SS}}(t) \nonumber\\
    &&= W^\dagger\left[\sum_a\dot\alpha_a\frac{\partial}{\partial \alpha_a}\right]\mathbf{p}_{\mathrm{SS}}(t) \nonumber\\
    &&=\sum_a\dot\alpha_a W^\dagger \frac{\partial \mathbf{p}_{\mathrm{SS}}}{\partial \alpha_a} \nonumber\\
    &&= \sum_a\dot\alpha_a  \mathbf{p}^{(1)}_{a} \quad \mathrm{where} \quad \mathbf{p}^{(1)}_{a} = W^\dagger \frac{\partial \mathbf{p}_{\mathrm{SS}}}{\partial \alpha_a}
\end{eqnarray}

2nd order:

\begin{eqnarray}
    \sum_{a,b} \dot\alpha_a  &&\dot \alpha_b  W^\dagger\left( \frac{\partial W^\dagger}{\partial \alpha_a} \frac{\partial \mathbf{p}_{\mathrm{SS}}}{\partial \alpha_b} + W^\dagger \frac{\partial^2 \mathbf{p}_{\mathrm{SS}}}{\partial \alpha_a \partial \alpha_b}  \right) \nonumber\\
    && \qquad + \sum_a \ddot \alpha_a (W^\dagger)^2 \frac{\partial \mathbf{p}_{\mathrm{SS}}}{\partial \alpha_a} \nonumber\\
    && =\sum_{a,b} \dot\alpha_a  \dot \alpha_b \mathbf{p}^{(2)}_{ab} +  \sum_a \ddot\alpha_a \mathbf{p}^{(3)}_a \nonumber\\
    &&\mathrm{where} \quad \mathbf{p}^{(2)}_{ab} = W^\dagger\left( \frac{\partial W^\dagger}{\partial \alpha_a} \frac{\partial \mathbf{p}_{\mathrm{SS}}}{\partial \alpha_b} + W^\dagger \frac{\partial^2 \mathbf{p}_{\mathrm{SS}}}{\partial \alpha_a \partial \alpha_b}  \right) \nonumber\\
    &&\mathrm{and} \quad  \mathbf{p}^{(3)}_{a} = (W^\dagger)^2 \frac{\partial \mathbf{p}_{\mathrm{SS}}}{\partial \alpha_a}
\end{eqnarray}

$\sigma$ is also expanded in several variables:

\begin{eqnarray}
	\sigma = \sigma&& \vert_{\mathbf{p} = \mathbf{p}_{\mathrm{SS}}} + \sum_a \epsilon_a \frac{\partial \sigma}{\partial \epsilon_a}\vert_{\mathbf{p} = \mathbf{p}_{\mathrm{SS}}} \nonumber\\
    &&+ \sum_{a,b}\epsilon_a\epsilon_b \frac{1}{2}\frac{\partial^2 \sigma}{\partial \epsilon_a\partial\epsilon_b}\vert_{\mathbf{p} = \mathbf{p}_{\mathrm{SS}}} + \mathcal{O}(\epsilon^3)
\end{eqnarray}

The entropy expansion is similar to the single-parameter case:\\

The 0th term:

\begin{eqnarray}
	\sigma \vert_{\mathbf{p} = \mathbf{p}_{\mathrm{SS}}}  = k_B \sum_{i,j,(\nu)} W_{ij}^{(\nu)}p_{\mathrm{SS}, j}\ln\frac{W_{ij}^{(\nu)}p_{\mathrm{SS}, j}}{W_{ji}^{(\nu)}p_{\mathrm{SS}, i}}
\end{eqnarray}

The 1st term:

\begin{eqnarray}
    \frac{\partial \sigma}{\partial \epsilon_a} &&\vert_{\mathbf{p} = \mathbf{p}_{\mathrm{SS}}} = k_B \sum_{i, j, (\nu)} W_{ij}^{(\nu)}\frac{\partial{p_j}}{\partial \epsilon_a}\ln\frac{W_{ij}^{(\nu)}p_{\mathrm{SS}, j}}{W_{ji}^{(\nu)}p_{\mathrm{SS}, i}} \nonumber\\ 
    &&= k_B \sum_{i, j, (\nu)} W_{ij}^{(\nu)}p^{(1)}_{a, j}\ln\frac{W_{ij}^{(\nu)}p_{\mathrm{SS}, j}}{W_{ji}^{(\nu)}p_{\mathrm{SS}, i}} \nonumber\\
    &&= k_B \sum_{i, j, k, (\nu)} \ln\frac{W_{ij}^{(\nu)}p_{\mathrm{SS}, j}}{W_{ji}^{(\nu)}p_{\mathrm{SS}, i}}  W_{ij}^{(\nu)}W^\dagger_{jk} \frac{\partial p_{\mathrm{SS}, k}}{\partial \alpha_a} \quad
\end{eqnarray}

\begin{widetext}

The 2nd term:

\begin{eqnarray}
    \frac{\partial^2 \sigma}{\partial \epsilon_a \partial \epsilon_b}&&\vert_{\mathbf{p} = \mathbf{p}_{\mathrm{SS}}}   = k_B \sum_{i,j,(\nu)}2\ W_{ij}^{(\nu)}\Bigg[ \frac{\partial ^2 p_j}{\partial \epsilon_a \partial \epsilon_b}\ln\frac{W_{ij}^{(\nu)}p_{\mathrm{SS}, j}}{W_{ji}^{(\nu)}p_{\mathrm{SS}, i}}-\frac{\partial p_i}{\partial \epsilon_a} p_j^{-1}\frac{\partial p_j}{\partial \epsilon_b}\Bigg]\nonumber\\
    &&= k_B \sum_{i,j,(\nu)}2\ W_{ij}^{(\nu)}\Bigg[p^{(2)}_{ab, j}\ln\frac{W_{ij}^{(\nu)}p_{\mathrm{SS}, j}}{W_{ji}^{(\nu)}p_{\mathrm{SS}, i}}-p^{(1)}_{a, i} p_j^{-1}p^{(1)}_{b, j}\Bigg] \quad 
    \left( + k_B \sum_{i,j,(\nu)}\ 2 W_{ij}^{(\nu)}p^{(3)}_{a,j}\ln\frac{W_{ij}^{(\nu)}p_{\mathrm{SS}, j}}{W_{ji}^{(\nu)}p_{\mathrm{SS}, i}} \quad \mathrm{if} \,\, a = b \right)\nonumber\\
    &&=k_B \sum_{i,j,k,l, (\nu)}\ 2W_{ij}^{(\nu)}\Bigg[\ln\frac{W_{ij}^{(\nu)}p_{\mathrm{SS}, j}}{W_{ji}^{(\nu)}p_{\mathrm{SS}, i}}W^\dagger_{jk}\left( \frac{\partial W^\dagger_{kl}}{\partial \alpha_a} \frac{\partial p_{\mathrm{SS}, l}}{\partial \alpha_b} + W^\dagger_{kl} \frac{\partial^2 p_{\mathrm{SS},l}}{\partial \alpha_a \partial \alpha_b}  \right) -W^\dagger_{ik} \frac{\partial p_{\mathrm{SS}, k}}{\partial \alpha_a} p_j^{-1}W^\dagger_{jl} \frac{\partial p_{\mathrm{SS}, l}}{\partial \alpha_b}\Bigg]\nonumber\\
    && \left( + k_B \sum_{i,j, k, l, (\nu)} 2W_{ij}^{(\nu)}\ln\frac{W_{ij}^{(\nu)}p_{\mathrm{SS}, j}}{W_{ji}^{(\nu)}p_{\mathrm{SS}, l}}\ W^\dagger_{jk}W^\dagger_{kl} \frac{\partial p_{\mathrm{SS},l}}{\partial \alpha_a} \quad \mathrm{if} \,\, a = b \right)
\end{eqnarray}

As in the 1-parameter case, we now fill in the full expression for each term in the expansion of $\sigma$. 

\begin{eqnarray}
	\sigma= k_B \sum_{i,j,(\nu)} && W_{ij}^{(\nu)}p_{\mathrm{SS}, j}\ln\frac{W_{ij}^{(\nu)}p_{\mathrm{SS}, j}}{W_{ji}^{(\nu)}p_{\mathrm{SS}, i}} + \sum_a \epsilon_a k_B \sum_{i, j, k, (\nu)} \ln\frac{W_{ij}^{(\nu)}p_{\mathrm{SS}, j}}{W_{ji}^{(\nu)}p_{\mathrm{SS}, i}}  W_{ij}^{(\nu)}W^\dagger_{jk} \frac{\partial p_{\mathrm{SS}}}{\partial \alpha_a} \nonumber\\
    &&+ \sum_{a,b}\epsilon_a\epsilon_bk_B \sum_{i,j,k,l, (\nu)}\ln\frac{W_{ij}^{(\nu)}p_{\mathrm{SS}, j}}{W_{ji}^{(\nu)}p_{\mathrm{SS}, i}} W_{ij}^{(\nu)} W^\dagger_{jk}\left( \frac{\partial W^\dagger_{kl}}{\partial \alpha_a} \frac{\partial p_{\mathrm{SS}, l}}{\partial \alpha_b} + W^\dagger_{kl} \frac{\partial^2 p_{\mathrm{SS},l}}{\partial \alpha_a \partial \alpha_b}  \right) \\ 
    &&- \sum_{a,b}\epsilon_a\epsilon_bk_B \sum_{i,j} \frac{\partial p_{\mathrm{SS}, i}}{\partial \alpha_a} W^\dagger_{ij}p_j^{-1} \frac{\partial p_{\mathrm{SS}, j}}{\partial \alpha_b}  + \sum_{a}\epsilon_a^2\ \sum_{i,j, k, l, (\nu)}\ln\frac{W_{ij}^{(\nu)}p_{\mathrm{SS}, j}}{W_{ji}^{(\nu)}p_{\mathrm{SS}, i}}\ W_{ij}^{(\nu)}W^\dagger_{jk}W^\dagger_{kl} \frac{\partial p_{\mathrm{SS},l}}{\partial \alpha_a} + \mathcal{O}(\epsilon^3) \nonumber
\end{eqnarray}

\end{widetext}

The entropy expansion is then given by: 

\begin{eqnarray}
	\sigma(t) = \sigma_ {SS}&& + \boldsymbol{\dot\alpha} (t) \cdot \boldsymbol{F} \left ( \alpha(t) \right) \nonumber\\
    &&+ \boldsymbol{\dot\alpha}(t)\left( A^{(1)} + A^{(2)} \right) \boldsymbol{\dot\alpha} (t) +  \boldsymbol{\ddot\alpha} (t) \cdot \boldsymbol{A^{(3)}} \qquad \quad\label{eq:appendix_expansion_short_form_multiple_par}
\end{eqnarray}

Where $\boldsymbol{F}$ and $\boldsymbol{A^{(3)}}$ are vectors and $A^{(1)}$ and $A^{(2)}$ are matrixes.

\begin{eqnarray}
	\sigma_{SS} &=& 
	k_B \sum_{i,j,(\nu)} \ln\frac{W_{ij}^{(\nu)}p_{\mathrm{SS}, j}}{W_{ji}^{(\nu)}p_{\mathrm{SS}, i}}W_{ij}^{(\nu)}\left( p_{\mathrm{SS}, j}  \right) \\
	F_a &=&  k_B \sum_{i, j, k, (\nu)} \ln\frac{W_{ij}^{(\nu)}p_{\mathrm{SS}, j}}{W_{ji}^{(\nu)}p_{\mathrm{SS}, i}}W_{ij}^{(\nu)} \left(W_{jk}^\dagger\frac{\partial}{\partial \alpha_a}p_{\mathrm{SS}, k} \right) \,\sim \boldsymbol{\dot\alpha}\nonumber\\\\
	A^{(1)}_{ab} &=& - k_B \sum_{i,j,(\nu)}\frac{\partial p_{\mathrm{SS}, i}}{\partial \alpha_a} p_{\mathrm{SS}, i}^{-1} W_{ij}^\dagger \frac{\partial p_{\mathrm{SS}, j}}{\partial \alpha_b} \, \sim   (\boldsymbol{\dot\alpha})^2 
\end{eqnarray}

\begin{eqnarray}
	A^{(2)}_{ab} &=& k_B \sum_{i,j,k, l(\nu)}
	\ln\frac{W_{ij}^{(\nu)}p_{\mathrm{SS}, j}}{W_{ji}^{(\nu)}p_{\mathrm{SS}, i}} W_{ij}^{(\nu)}\nonumber\\
    &&\qquad\times\left( W_{jk}^\dagger\frac{\partial}{\partial \alpha_a}\left( W_{kl}^\dagger \frac{\partial p_{\mathrm{SS}, l}}{\partial \alpha_b} \right) \right) \,\sim (\boldsymbol{\dot\alpha})^2\nonumber\\
    \\
	A^{(3)}_a &=& k_B \sum_{i,j,k, l (\nu)}
	\ln\frac{W_{ij}^{(\nu)}p_{\mathrm{SS}, j}}{W_{ji}^{(\nu)}p_{\mathrm{SS}, i}}W_{ij}^{(\nu)}\nonumber\\
    && \qquad \times\left( W^\dagger_{jk}W^\dagger_{kl}\frac{\partial p_{\mathrm{SS}, l}}{\partial \alpha_a} \right) \,\sim \boldsymbol{\ddot\alpha}
\end{eqnarray}
    The $A^{(3)}$-term can be reformulated to depend on  $\left(\dot\alpha\right)^2$ instead of $\ddot\alpha$:

\begin{eqnarray}
	\ddot\alpha (t)  A^{(3)}= \frac{\partial}{\partial \alpha} \left(\dot\alpha A^{(3)}\right)-(\dot\alpha)^2\frac{\partial}{\partial \alpha}A^{(3)}
\end{eqnarray}

The term $\partial/\partial \alpha\left(\dot\alpha A^{(3)}\right)$ is evaluated at the boundaries in $S_{tot}$ and depends only on initial values of the protocol:

\begin{eqnarray}
	\Delta &&S_{\mathrm{tot}} = \int_{t_i}^{t_f} \sigma(t) dt \nonumber\\
    &&= \int_{t_i}^{t_f}\sigma_ {SS}(t)  dt + \int_{\alpha_0}^ {\alpha_f}  F \left ( \alpha(t) \right)  \cdot  d\alpha \nonumber\\
	&& + \int_{t_i}^{t_f} \dot\alpha (t) \left( A^{(1)} (\alpha (t)) + A^{(2)} (\alpha (t)) - A^{(3)}(\alpha (t)) \right) \dot\alpha (t)  dt \nonumber\\
    &&+  \left[ \dot\alpha A^{(3)} \right]_{t_i}^{t_f}
\end{eqnarray}

This simplifies calculations and allows for optimization using Euler-Lagrange.

    \section{\label{app:rewrite_W_dagger} Rewriting $W^\dagger$ as an time-integral in steady-state} 
    $W^\dagger$ can be found by integrating the Green's function for the system over all time at steady state (holding $\vert \mathbf{p}_{SS} \rangle$ and $W$ constant), thereby sampling all trajectories/transitions:

\begin{eqnarray}
    W^\dagger = -\int_0^\infty \exp(W \tau) \left( \mathrm{Id} - \vert \mathbf{p}_{SS} \rangle \langle 1 \vert \right) d\tau
    \label{eq:appendix_w_dagger_int}
\end{eqnarray}

Here $ \left( \mathrm{Id} - \vert \mathbf{p}_{SS} \rangle \langle 1 \vert \right)$ makes sure that any components of the first set of eigenvectors are not included. Furthermore, we have: 

\begin{eqnarray}
\left(W^\dagger \right)^2 = \int_0^\infty \exp(W \tau) \tau \left( \mathrm{Id} - \vert \mathbf{p}_{SS} \rangle \langle 1 \vert \right) d\tau
\label{eq:appendix_w_dagger_sq_int}
\end{eqnarray}

This is seen by partial integration. First we note that the Green's function, $G$, for the system denotes the solution to the master equation:

\begin{eqnarray}
   \frac{d}{dt}\mathbf{p}_x(t) = \sum_{x'} W_{xx'} \mathbf{p}_{x'}(t')
\end{eqnarray}

That can take the probability distribution at a time $t'$ and translate it to $t$:

\begin{eqnarray}
   \mathbf{p}_x(t) = \sum_{x'} G_{xx'}(t, t') \mathbf{p}_{x'}(t')
\end{eqnarray}

And is given by solving the 1st order differential equation, with the initial condition that with no time passing, the probability distribution should be unaltered ($G_{xx'}(t', t')  = \delta_{xx'}^K$) \cite{peliti_stochastic_2021}:

\begin{eqnarray}
  G_{xx'}(t, t') = \exp \left( \int_{t'}^tW(t'')dt''\right)_{xx'}
\end{eqnarray}

In this case, $W$ is constant and $t' = 0$, and we have:

\begin{eqnarray}
  G_{xx'}(t, 0) = \exp \left( \int_{0}^tWdt''\right)_{xx'} =  \exp \left(W t\right)_{xx'} \quad
\end{eqnarray}

We note that for a matrix, $W$, with the eigenvalues $\{\lambda_i\}_{i \in \mathbb{N}^+}$, and a constant, $k$, the matrix $kW$ share the same eigenvectors with $W$, and the eigenvalues are given by $\{k\lambda_i\}_{i \in \mathbb{N}^+}$. Since $W$ is symmetric and therefore diagonizable, the matrix exponential can then be expressed as:

\begin{eqnarray}
\exp(W \tau) =
\sum_{i = 0}^n \vert \lambda_i \rangle \langle \lambda_i \vert \exp(\lambda_i \tau) 
\label{eq:appendix_matrix_exp}
\end{eqnarray}

We insert eq.~\ref{eq:appendix_matrix_exp} into \ref{eq:appendix_w_dagger_int} and expand the equation using integration by parts:

\begin{eqnarray}
-\int_0^\infty && \exp(W \tau) \left( \mathrm{Id} - \vert \mathbf{p}_{SS} \rangle \langle 1 \vert \right) d\tau \nonumber\\
&&= - \int_0^\infty \sum_{i = 0}^n \exp(\lambda_i \tau) \vert \lambda_i \rangle \langle \lambda_i \vert  \left( \mathrm{Id} - \vert \mathbf{p}_{SS} \rangle \langle 1 \vert \right) d\tau \nonumber\\
&&= -\left[\sum_{i = 1}^n 1 \cdot \exp(\lambda_i \tau) \cdot  \lambda_i^{-1}\vert \lambda_i \rangle \langle \lambda_i \vert \right]_0^\infty\nonumber\\ 
&& \quad + \int_0^\infty \sum_{i = 1}^n 1 \cdot \exp(\lambda_i \tau) \cdot \lambda_i^{-1}\vert \lambda_i \rangle \langle \lambda_i \vert \cdot 0 \cdot d\tau \nonumber\\
&&= \sum_{i = 1}^n \lambda_i^{-1}\vert \lambda_i \rangle \langle \lambda_i \vert \quad \text{assuming $\lambda_i < 0$ for all $i$}\nonumber\\
&&= W^\dagger
\end{eqnarray}

And likewise filling eq.~\ref{eq:appendix_matrix_exp} into \ref{eq:appendix_w_dagger_sq_int} and expand the equation using integration by parts: 

\begin{eqnarray}
\int_0^\infty &&\exp(W \tau) \tau \left( \mathrm{Id} - \vert \mathbf{p}_{SS} \rangle \langle 1 \vert \right) d\tau \nonumber\\
&&= \int_0^\infty \sum_{i = 0}^n \tau  \exp(\lambda_i \tau) \vert \lambda_i \rangle \langle \lambda_i \vert\left( \mathrm{Id} - \vert \mathbf{p}_{SS} \rangle \langle 1 \vert \right) d\tau \nonumber\\
&&= \left[\sum_{i = 1}^{n} \tau  \exp(\lambda_i \tau) \lambda_i^{-1}\vert \lambda_i \rangle \langle \lambda_i \vert\ \right]_0^\infty \nonumber\\
&& \quad  -\int_0^\infty \sum_{i = 1}^{n} \exp(\lambda_i \tau) \lambda_i^{-1} \vert \lambda_i \rangle \langle \lambda_i \vert d\tau \nonumber\\
&& = - \left[ \sum_{i = 1}^n \exp(\lambda_i \tau) \lambda_i^{-2} \vert \lambda_i \rangle \langle \lambda_i \vert\ \right]_0^\infty \nonumber \\ && \qquad \qquad \qquad \qquad \qquad \qquad \text{(assuming $\lambda_i < 0$ for all $i$)} \nonumber\\
&&= \left( W^\dagger \right)^2 
\end{eqnarray}

    \section{\label{app:observables} Rewriting expansion terms using observables} 
    The second order expansion obtained in \ref{app:expansion_terms} can be rephrased as a function of observables for practical application.\\

If one defines the following 3 observables as in the main text:
\begin{eqnarray}
    O_{1; j} &=&  \sum_{i, (\nu)} W_{ij}^{(\nu)} \ln \frac{W_{ij}^{(\nu)}p_{SS, j}}{W_{ji}^{(\nu)}p_{SS, i}}
    \label{eq:appendix_observable_1}\\
    O_{2; a;j} &=&  \frac{\partial \ln p_{SS, j}}{\partial \alpha_a}
    \label{eq:appendix_observable_2}\\
    O_{3; a;ij} &=&  \delta(t-\tau_{ij})\frac{\partial \ln(W_{ij})}{\partial \alpha_a} \; \text{(elementwise)}
    \label{eq:appendix_observable_3}
\end{eqnarray}

The steady state dissipation, $\sigma_{SS}$, can be expressed by the first observable, \ref{eq:appendix_observable_1}:

\begin{eqnarray}
    \sigma^{SS} &&= \sum_{i,j, (\nu)} W_{ij}^{(\nu)} p_{SS, j} \ln \frac{W_{ij}^{(\nu)}p_{SS, j} }{W_{ji}^{(\nu)}p_{SS, i}} \nonumber\\
    &&= \sum_j p_{SS, j}  \sum_{i, (\nu)} W_{ij}^{(\nu)} \ln \frac{W_{ij}^{(\nu)}p_{SS, j} }{W_{ji}^{(\nu)}p_{SS, i}} \nonumber\\
    &&= \langle O_1\rangle
\end{eqnarray}

$F_a$ is expressed as a combination of observable 1 and 2:

\begin{eqnarray}
     F_a &&= \sum_{i, j, k, (\nu)} \ln \frac{W_{ij}^{(\nu)}p_{SS, j} }{W_{ji}^{(\nu)}p_{SS, i}} W_{ij}^{(\nu)} W_{jk}^\dagger \frac{\partial p_{SS, k}}{\partial \alpha_a} \nonumber\\
    &&= -\sum_{i, j, k, (\nu)} \left( \ln \frac{W_{ij}^{(\nu)}p_{SS, j} }{W_{ji}^{(\nu)}p_{SS, i}} W_{ij}^{(\nu)} \right) \nonumber\\
    &&\qquad \quad\times\left( \int_0^\infty\exp(W\tau)\left( \mathrm{Id} - \vert p_{SS} \rangle \langle 1 \vert \right) d\tau \right)_{jk} p_{SS, k}\frac{\partial p_{SS, k}}{\partial \alpha_a} \nonumber\\
    &&=  -\int_0^\infty \sum_k p_{SS, k} \sum_{j, (\nu)}  O_{1, j}(\tau) \nonumber\\
    &&\qquad \qquad \times\left( \exp(W\tau)\left( \mathrm{Id} - \vert p_{SS} \rangle \langle 1 \vert \right)\right)_{jk} O_{2, a, k}(\tau)d\tau
\end{eqnarray}

The Green's function propagates the first observable back in time:

\begin{eqnarray}
 \left(\exp(W\tau)\left( \mathrm{Id} - \vert \mathbf{p_{SS}} \rangle \langle 1 \vert \right)\right)_{jk} &&O_{2, a, k}(\tau) \nonumber\\
 &&=  O_{2, a, k}(0)
\end{eqnarray}

Continuing from previously:

\begin{eqnarray}
    F_a&&= -\int_0^\infty \sum_{j, (\nu)}  \langle O_{1, j} \exp(W\tau)\left( \mathrm{Id} - \vert \mathbf{p}_{SS} \rangle \langle 1 \vert \right)_{j,k} O_{2, a, k} \rangle_k d\tau \nonumber \\
    &&= -\int_0^\infty \langle O_{1} (t)  O_{2, a} (0) \rangle d\tau
\end{eqnarray}

$A^{(1)}_{ab}$ is solely depending on observable 2:

\begin{eqnarray}
     A^{(1)}_{ab} &&= \sum_{i, j}p_{SS, j}\frac{\partial \ln(p_{SS, i})}{\partial \alpha_b}W_{ij}^\dagger\frac{\partial \ln(p_{SS, j})}{\partial \alpha_a} \nonumber\\
    &&= -\sum_{j}p_{SS, j} \sum_{i}\frac{\partial \ln(p_{SS, i})}{\partial \alpha_b} \nonumber\\
    && \qquad \quad \times\left( \int_0^\infty\exp(W\tau)\left( \mathrm{Id} - \vert \mathbf{p}_{SS} \rangle \langle 1 \vert \right) d\tau \right)_{ij}\frac{\partial \ln(p_{SS, j})}{\partial \alpha_a} \nonumber\\
    &&= -\int_0^\infty \sum_{i} \langle \frac{\partial \ln(p_{SS, i})}{\partial \alpha_b} \nonumber\\
    && \qquad \quad\times \left( \exp(W\tau)\left( \mathrm{Id} - \vert \mathbf{p}_{SS} \rangle \langle 1 \vert \right)\right)_{ij}\frac{\partial \ln(p_{SS, j})}{\partial \alpha_a} \rangle_j d\tau \nonumber\\
    &&= -\int_0^\infty \langle O_{2, b} (\tau)  O_{2, a} (0) \rangle d\tau
\end{eqnarray}

\begin{widetext}

$A^{(2)}_{a}$ is a little more elaborate. First we expand:

\begin{eqnarray}
     A_{ab}^{(2)} &&= \sum_{i,j,k,l, (\nu)} \ln\left(\frac{W_{ij}^{(\nu)}p_{SS, j}}{W_{ji}^{(\nu)}p_{SS, i}}\right)W_{ij}^{(\nu)}W^\dagger_{jk} \frac{\partial }{\partial \alpha_a} \left(W_{kl}^{\dagger} \frac{\partial p_{SS, l}}{\partial \alpha_b} \right) \nonumber\\
    &&= \sum_{i,j,k,l, (\nu)} \ln\left(\frac{W_{ij}^{(\nu)}p_{SS, j}}{W_{ji}^{(\nu)}p_{SS, i}}\right)W_{ij}^{(\nu)}W^\dagger_{jk} \left( \frac{\partial W_{kl}^{\dagger}}{\partial \alpha_a} \frac{\partial p_{SS, l}}{\partial \alpha_b} + W_{kl}^{\dagger} \frac{\partial p_{SS, l}}{\partial \alpha_a \partial \alpha_b} \right) \nonumber\\
    &&= \sum_{i,j,k,l, (\nu)} \ln\left(\frac{W_{ij}^{(\nu)}p_{SS, j}}{W_{ji}^{(\nu)}p_{SS, i}}\right)W_{ij}^{(\nu)}W^\dagger_{jk} \left( \frac{\partial W_{kl}^{\dagger}}{\partial \alpha_a} \frac{\partial \ln (p_{SS, l})}{\partial \alpha_b}p_{SS,l} + W_{kl}^{\dagger} \frac{\partial }{\partial \alpha_a} \left( \frac{\partial \ln(p_{SS, l})}{\partial \alpha_b} p_{SS, l}\right) \right)\nonumber\\
    &&= \sum_{i,j,k,l, (\nu)} \ln\left(\frac{W_{ij}^{(\nu)}p_{SS, j}}{W_{ji}^{(\nu)}p_{SS, i}}\right)W_{ij}^{(\nu)}W^\dagger_{jk} \nonumber\\
    && \quad \quad \quad \cdot \left( \frac{\partial W_{kl}^{\dagger}}{\partial \alpha_a} \frac{\partial \ln (p_{SS, l})}{\partial \alpha_b}p_{SS,l} + W_{kl}^{\dagger} \left( \frac{\partial^2 \ln(p_{SS, l})} {\partial \alpha_a \partial \alpha_b} p_{SS, l} + \frac{\partial \ln(p_{SS, l})}{\partial \alpha_b} \frac{\partial \ln(p_{SS, l})}{\partial \alpha_a}p_{SS, l}\right) \right) \nonumber\\
    &&= \sum_{j,k, l (\nu)} O_{1;j} W^\dagger_{jk}
    \left(\frac{\partial W_{kl}^{\dagger}}{\partial \alpha_a} O_{2;b;l}  + W_{kl}^{\dagger} \left( \frac{\partial O_{2;b;l}} {\partial \alpha_a} +O_{2;b;l} O_{2;a;l} \right) \right) p_{SS,l}\nonumber\\
    &&= \sum_{j,k, l, (\nu)} O_{1;j} W^\dagger_{jk} \frac{\partial W_{kl}^{\dagger}}{\partial \alpha_a} O_{2;b;l}  p_{SS, l}+ \sum_{j,k,l, (\nu)} O_{1;j} W^\dagger_{jk} W_{kl}^{\dagger} \left( \frac{\partial O_{2;b;l}} {\partial \alpha_a} +O_{2;b;l} O_{2;a;l} \right) p_{SS,l}
    \label{eq:appendix_A2_term_0}
\end{eqnarray}

\end{widetext}

The latter term of \ref{eq:appendix_A2_term_0} can be rewritten quite easily, using the same argument as in earlier:

\begin{eqnarray}
     \sum_{j,k,l, (\nu)}&& O_{1;j} W^\dagger_{jk} W_{kl}^{\dagger} \left( \frac{\partial O_{2;b;l}} {\partial \alpha_a} +O_{2;b;l} O_{2;a;l} \right) p_{SS,l} \nonumber\\
     &&= \int_0^\infty \sum_{j,k, l, (\nu)}  O_{1;j} \left( \exp(W \tau)\left( \mathrm{Id} - \vert \mathbf{p}_{SS} \rangle \langle 1 \vert \right)\right)_{jl} \nonumber\\
     && \qquad \qquad \times\tau \left( \frac{\partial O_{2;b;l}} {\partial \alpha_a} +O_{2;b;l} O_{2;a;l} \right) p_{SS, l}d\tau\nonumber\\
     && = \int_0^\infty \langle O_{1}(\tau) \tau \left( \frac{\partial O_{2;b}} {\partial \alpha_a} \vert_{\tau = 0}+O_{2;b}(0) O_{2;a}(0) \right) \rangle d\tau \nonumber\\
     &&\label{eq:appendix_A2_term_2}
\end{eqnarray}

For the first term of \ref{eq:appendix_A2_term_0}, we need to rewrite the term $\partial W^{\dagger}/\partial \alpha_a$. First we utilize that $W^\dagger$ can be written as:

\begin{eqnarray}
W^\dagger &&\left( W^\dagger W \right) \nonumber\\
&&= W^\dagger \left( \mathrm{Id} - \vert \mathbf{p}_{SS} \rangle \langle 1\vert \right) \nonumber\\
&&= W^\dagger - 0 = W^\dagger
\end{eqnarray}

Thus $W^\dagger\left(\partial W^{\dagger}/\partial \alpha_a \right)$ can be rewritten as:

\begin{eqnarray}
    W^\dagger\frac{\partial W^{\dagger}}{\partial \alpha_a} &&= W^\dagger W^\dagger W\frac{\partial W^{\dagger}}{\partial \alpha_a} \nonumber\\
    &&= (W^\dagger)^2 \left( \frac{\partial}{\partial \alpha_a} \left[WW^\dagger\right] - \frac{\partial W}{\partial \alpha_a} W^\dagger\right)\nonumber\\
    &&= (W^\dagger)^2 \left( \frac{\partial}{\partial \alpha_a} \left[ \mathrm{Id} - \vert \mathbf{p}_{SS} \rangle \langle 1 \vert \right] - \frac{\partial W}{\partial \alpha_a} W^\dagger \right) \nonumber\\
    &&= (W^\dagger)^2 \left(  - \frac{\partial}{\partial \alpha_a} \vert \mathbf{p}_{SS} \rangle \langle 1 \vert - \frac{\partial W}{\partial \alpha_a} W^\dagger \right)
    \label{eq:appendix_observables_Einstein_notation}
\end{eqnarray}

When formulating this in Einstein notation, it is noted that the second term in parentheses in \ref{eq:appendix_observables_Einstein_notation} requires an additional index for summation compared to the first one. This is solved by multiplying the first term with a delta-function on the index:

\begin{eqnarray}
    (W^\dagger)^2_{jk} \left(  - \frac{\partial  p_{SS, k}}{\partial \alpha_a} \delta_{lm}  - \frac{\partial W_{kl}}{\partial \alpha_a} W^\dagger_{lm} \right)
\end{eqnarray}

We plug this into the first term of \ref{eq:appendix_A2_term_0}:

\begin{eqnarray}
     \sum_{j,k, l, (\nu)} &&O_{1;j} W^\dagger_{jk} \frac{\partial W_{kl}^{\dagger}}{\partial \alpha_a} O_{2;b;l}  p_{SS, l} \\
     &&= \sum_{j,k, l, m (\nu)}  O_{1;j} (W^\dagger)^2_{jk} \nonumber \\
     && \qquad \quad \times\left(  - \frac{\partial  p_{SS, k}}{\partial \alpha_a} \delta_{lm}  - \frac{\partial W_{kl}}{\partial \alpha_a} W^\dagger_{lm} \right) O_{2;b;m} p_{SS, m} \nonumber
\end{eqnarray}

We split up the integral:

\begin{eqnarray}
    \sum_{j,k, l, m (\nu)}  &&O_{1;j} (W^\dagger)^2_{jk} \left(  - \frac{\partial  p_{SS, k}}{\partial \alpha_a} \delta_{lm}  - \frac{\partial W_{kl}}{\partial \alpha_a} W^\dagger_{lm} \right) O_{2;b;m} p_{SS, m} \nonumber\\
     &&= -\sum_{j,k, l, m (\nu)}  O_{1;j} (W^\dagger)^2_{jk} \frac{\partial  p_{SS, k}}{\partial \alpha_a} \delta_{lm} O_{2;b;m} p_{SS, m} \nonumber\\
     && \quad \quad  - \sum_{j,k, l, m (\nu)}  O_{1;j} (W^\dagger)^2_{jk} \frac{\partial W_{kl}}{\partial \alpha_a} W^\dagger_{lm} O_{2;b;m} p_{SS, m}\nonumber\\
\end{eqnarray}

And rewrite  $\left(W^\dagger\right)^2$ as a time-integral, $ \left(\partial W/\partial \alpha_a \right) W^\dagger$ as $ \left(\partial \ln(W)/\partial \alpha_a \right) \left( \mathrm{Id} - \vert \mathbf{p}_{SS} \rangle \langle 1 \vert \right)$ and $\partial p_{SS}/\partial \alpha_a$ as observable 2 (see \ref{eq:appendix_observable_2}):

\begin{eqnarray}
     = -&&\int_0^{\infty}\sum_{j,k, l, m (\nu)}  O_{1;j}\left( \exp(W \tau)\tau\left( \mathrm{Id} - \vert \mathbf{p}_{SS} \rangle \langle 1 \vert \right)\right)_{jk} \nonumber\\\nonumber\\
     &&  \times O_{2;a;k} p_{SS, k} \delta_{lm} O_{2;b;m} p_{SS, m} d\tau \nonumber\\\nonumber\\
     &&- \int_0^{\infty}\int_0^{\infty}\sum_{j,k, l, m (\nu)}  O_{1;j} \left( \exp(W \tau_1)\tau_1\left( \mathrm{Id} - \vert \mathbf{p}_{SS} \rangle \langle 1 \vert \right)\right)_{jk} \nonumber\\
     && \qquad \qquad \times \,\frac{\partial W_{kl}}{\partial \alpha_a} \left( \exp(W \tau_2)\left( \mathrm{Id} - \vert \mathbf{p}_{SS} \rangle \langle 1 \vert \right)\right)_{lm} \nonumber \\
     && \qquad \qquad \qquad \qquad \times O_{2;b;m} p_{SS, m} d\tau_1 d\tau_2
    \label{eq:appendix_A2_term_1}
\end{eqnarray}

$O_{2;a}(\tau)$ is propagated by the Greens function from $t = \tau$ to $t = 0$:
\begin{eqnarray}
\sum_k\left( \exp(W \tau)\left( \mathrm{Id} - \vert \mathbf{p}_{SS} \rangle \langle 1 \vert \right)\right)_{jk}&&O_{2;a;k}(\tau) \nonumber\\
&&= O_{2;a;j}(0)
\end{eqnarray}

\noindent And the first half of \ref{eq:appendix_A2_term_1} is now expressed as correlation functions between the observables $O_{1}$ and $O_{2}$:
\begin{eqnarray}
-\int_0^{\infty}&&\sum_{j,k, l, m (\nu)}  O_{1;j} \left( \exp(W \tau)\tau\left( \mathrm{Id} - \vert \mathbf{p}_{SS} \rangle \langle 1 \vert \right)\right)_{jk} \nonumber\\
&& \qquad \qquad \qquad \times O_{2;a;k} p_{SS, k} \delta_{lm} O_{2;b;m} p_{SS, m} d\tau \nonumber\\
&&=-\int_0^{\infty}\sum_{j, (\nu)} \langle O_{1;j}(\tau) \left( \exp(W \tau)\tau\left( \mathrm{Id} - \vert \mathbf{p}_{SS} \rangle \langle 1 \vert \right)\right)_{jk}  \nonumber\\
&& \qquad \qquad \qquad \times O_{2;a;k}(\tau) \rangle_k  \langle O_{2;b;m} \rangle_m d\tau \nonumber\\
&&= -\int_0^{\infty} \langle O_{1}(\tau) \tau O_{2;a}(0)\rangle \langle O_{2;b} \rangle d\tau
\label{eq:appendix_A2_term_1_1}
\end{eqnarray}

$\partial W/\partial \alpha_a$ in the last term of \ref{eq:appendix_A2_term_1} is substituted for the observable $O_{3; a; kl}$ by doing the following: Since none of the entries in $W$ are zero, we can transform the variable:

\begin{eqnarray}
\frac{\partial W_{kl}}{\partial \alpha_a} = W_{kl}\frac{\partial \ln (W_{kl})}{\partial \alpha_a}
\end{eqnarray}

The ensemble average of this quantity is given by:

\begin{eqnarray}
\sum_{l} \frac{\partial \ln (W_{kl})}{\partial \alpha_a}W_{kl} p_{SS, l} 
\end{eqnarray}

This can be seen as the average of $\partial \ln (W_{kl})/\partial \alpha_a$ over the probability flux into state $k$. \\


Integrating this quantity over all time at steady state:

\begin{eqnarray}
\int_0^{\infty}\sum_{l} \frac{\partial \ln (W_{kl})}{\partial \alpha_a}W_{kl} p_{SS, l} dt
\end{eqnarray}

is synonymous with integrating the ensemble average for $\partial \ln (W_{kl})/\partial \alpha_a$ over all time,

\begin{eqnarray}
\sum_{l} \frac{\partial \ln (W_{kl})}{\partial \alpha_a}p_{SS, l}
\end{eqnarray}

but sampling only when a transition from state $l$ to state $k$ occurs:

\begin{eqnarray}
\int_0^\infty \sum_{l} \delta(t - \tau_{kl})\frac{\partial \ln (W_{kl})}{\partial \alpha_a}p_{SS, l} dt
\end{eqnarray}

\begin{widetext}
\newpage 

The latter term of \ref{eq:appendix_A2_term_1} can then be rewitten as: 

\begin{eqnarray}
- \int_0^{\infty}&&\int_0^{\infty}\sum_{j,k, l, m, (\nu)}  O_{1;j} \left( \exp(W \tau_1)\tau_1\left( \mathrm{Id} - \vert \mathbf{p}_{SS} \rangle \langle 1 \vert \right)\right)_{jk} \frac{\partial W_{kl}}{\partial \alpha_a}\left( \exp(W \tau_2) \left( \mathrm{Id} - \vert \mathbf{p}_{SS} \rangle \langle 1 \vert \right)\right)_{lm} \nonumber\\
&& \qquad  \qquad  \qquad  \qquad  \qquad  \qquad  \qquad  \qquad  \qquad  \qquad  \qquad  \qquad  \qquad \times O_{2;b;m} p_{SS, m}  d\tau_1 d\tau_2 \nonumber\\
&& =- \int_0^{\infty}\int_0^{\infty}\sum_{j, l (\nu)} \langle O_{1;j}\tau_1\frac{\partial W_{jl}}{\partial \alpha_a}\vert_{t = - \tau_1} O_{2;b;l}(-(\tau_1 + \tau_2)) \rangle d\tau_1 d\tau_2 \nonumber\\
&& = - \int_0^{\infty}\int_0^{\infty}\sum_{j, l,(\nu)}  \langle O_{1;j}(\tau_1 + \tau_2) \tau_1  \frac{\partial W_{jl}}{\partial \alpha_a}\vert_{t = \tau_2} O_{2;b;l}(0) \rangle d\tau_1 d\tau_2\nonumber\\
&& = - \int_0^{\infty}\int_0^{\infty}\sum_{j, l,(\nu)}  \langle O_{1; j}(\tau_1 + \tau_2) \tau_1 O_{3;a, jl}(\tau_2) O_{2;b;l}(0) \rangle d\tau_1 d\tau_2 \nonumber\\
&& = - \int_0^{\infty}\int_0^{\infty}  \langle O_{1}(\tau_1 + \tau_2) \tau_1 O_{3;a}(\tau_2) O_{2;b}(0) \rangle d\tau_1 d\tau_2
\label{eq:appendix_A2_term_1_2}
\end{eqnarray}
\end{widetext}

Finally, one arrives at the following expression for $A_{ab}^{(2)}$ by adding \ref{eq:appendix_A2_term_1_1}, \ref{eq:appendix_A2_term_1_2}, and \ref{eq:appendix_A2_term_2}:

\begin{eqnarray}
     A_{ab}^{(2)} = &&-\int_0^{\infty} \langle O_{1}(\tau) \tau O_{2;a}(0)\rangle \langle O_{2;b} \rangle d\tau\\
     &&-\int_0^{\infty}\int_0^{\infty} \langle O_{1}(\tau_1 + \tau_2) \tau_1 O_{3;a}(\tau_2) O_{2;b}(0) \rangle d\tau_1 d\tau_2 \nonumber\\
    &&+\int_0^\infty \langle O_{1}(\tau) \tau \left( \frac{\partial O_{2;b}} {\partial \alpha_a} \vert_{\tau = 0}+O_{2;b}(0) O_{2;a}(0) \right) \rangle d\tau \nonumber
\end{eqnarray}

The $A_a^{(3)}$ term is as follows:

\begin{eqnarray}
    &&A^{(3)}_a = k_B \sum_{i,j,k, l (\nu)}
	\ln\frac{W_{ij}^{(\nu)}p_{\mathrm{SS}, j}}{W_{ji}^{(\nu)}p_{\mathrm{SS}, i}}W_{ij}^{(\nu)}\left( W^\dagger_{jk}W^\dagger_{kl}\frac{\partial p_{\mathrm{SS}, l}}{\partial \alpha_a} \right)\nonumber\\
    &&\qquad = k_B \sum_{j, l}
	\sum_{i,(\nu)} W_{ij}^{(\nu)} \ln\frac{W_{ij}^{(\nu)}p_{\mathrm{SS}, j}}{W_{ji}^{(\nu)}p_{\mathrm{SS}, i}} \nonumber\\
    &&\times\left( \left(\int_0^\infty \exp(W \tau) \tau \left( \mathrm{Id} - \vert \mathbf{p}_{SS} \rangle \langle 1 \vert \right) d\tau \right)_{jl}\frac{\partial \ln( p_{\mathrm{SS}, l})}{\partial \alpha_a}p_{\mathrm{SS}, l}\right)\nonumber\\
    &&\qquad = \int_0^{\infty}\langle O_1(\tau) \tau O_{2;a}(0)\rangle d\tau
\end{eqnarray}

    \section{\label{app:expansion_deviation} On the discrepancy between integration of the master equation and the expansion}
\begin{figure}[b]
    \centering
    \includegraphics[width=0.9\linewidth]{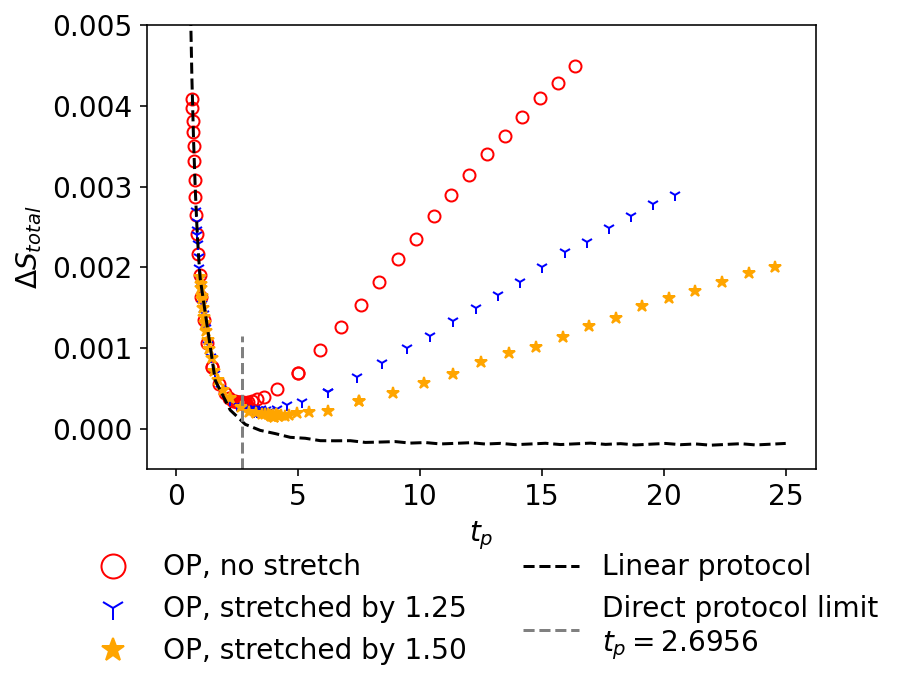}
    \caption{\textbf{The expansion deviates from the numerical integration.} Optimal protocols are marked 'OP'. Here we see the error increasing at longer protocol durations, $t_p$. The error is here the difference between the entropy production obtained by expansion and numerical integration of the master equation. Stretching the protocol, i.e. preserving the scale of the protocol but uniformly prolonging its duration, decreases the slope at which the error increases.}
    \label{fig:error_and_stretching}
\end{figure}

    For the linear protocols, the difference between the integrated expansion and the integrated master equation decreases as the duration of the protocol increases and the protocol slows down (see fig. \ref{fig:error_and_stretching}). \\

Oppositely, we find that the optimized protocols deviate from the numerical integration of the master equation at long protocol durations, see \ref{fig:error_and_stretching}. In contrast to the near-equilibrium picture, increasing the protocol duration does not entail $\dot \alpha \rightarrow 0$ for the optimal protocol and therefore the approximation might not always converge, even in the long duration limit. Longer protocol durations leads to increasingly extreme overshoots and the average energy of the system diverges, $E\rightarrow \infty$.  Stretching the protocols, i.e. decreasing $\dot\alpha$ but preserving the range of $\alpha$, counteracts this deviation, indicating that we are not within the $\dot\alpha$ range where a 2nd order expansion is a sufficient approximation for the dissipation.\\

\begin{widetext}
    \section{\label{app:terms_plots} Individual contribution of expansion terms versus protocol duration}
    \begin{figure}[ht]
    \centering
    \includegraphics[width=0.5\linewidth]{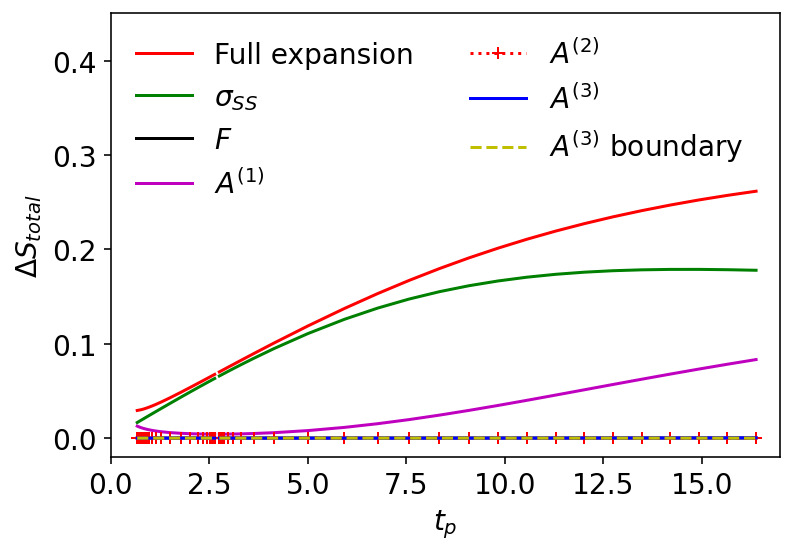}
    \caption{\textbf{Each term's contribution to the total entropy produced during an optimal protocol of duration $\boldsymbol{t_p}$.} Here we see that $\sigma_{\text{SS}}$ is the major contributor to the entropy production in this $t_p$ range, followed by $A^{(1)}$.}
    \label{fig:terms_plot_large_scale}
\end{figure}

\end{widetext}

\bibliography{references}

\end{document}